\documentclass[usegraphicx,usenatbib,usedcolumn]{mn2e}
\usepackage{longtable}

\title[A population synthesis study of PCEBs]{A comprehensive population
synthesis study of post-common envelope binaries}

\author[P. J. Davis et al.]{P.~J.~Davis,$^1$ U.~Kolb,$^1$ B.~Willems$^2$
\\ $^1$The Open University, Department of
 Physics and Astronomy, Walton Hall, Milton Keynes MK7 6AA \\
 $^2$Northwestern University, Department of Physics and Astronomy,
 2131 Tech Drive, Evanston, IL 60208, USA}
\begin{document}

\maketitle

\begin{abstract}
We apply population synthesis techniques to calculate the present
day population of post-common envelope binaries (PCEBs) for a range of
theoretical models describing the common envelope (CE) phase.
Adopting the canonical energy budget approach we consider models where
the ejection efficiency, $\alpha_{\rmn{CE}}$ is either a constant, or
a function of the secondary mass. We obtain the envelope binding
energy from detailed stellar models of the progenitor primary, with
and without the thermal and ionization energy, but we also test a
commonly used analytical scaling. We also employ the alternative
angular momentum budget approach, known as the $\gamma$-algorithm. We
find that a constant, global value of $\alpha_{\rmn{CE}} \ga 0.1$ can
adequately account for the observed population of PCEBs with late
spectral-type secondaries. However, this prescription fails to
reproduce IK Pegasi, which has a secondary with spectral type A8. We
can account for IK Pegasi if we include thermal and ionization energy
of the giant's envelope, or if we use the $\gamma$-algorithm. However,
the $\gamma$-algorithm predicts local space densities that are 1 to 2
orders of magnitude greater than estimates from observations. In
contrast, the canonical energy budget prescription with an initial
mass ratio distribution that favours unequal initial mass ratios
($n(q_{\rmn{i}})\propto{q_{\rmn{i}}^{-0.99}}$) gives a local space
density which is in good agreement with observations, and best
reproduces the observed distribution of PCEBs. Finally, all models
fail to reproduce the sharp decline for orbital periods,
$P_{\rmn{orb}} \ga 1$ d in the orbital period distribution of observed
PCEBs, even if we take into account selection effects against systems
with long orbital periods and early spectral-type secondaries.

\end{abstract}


\section{Introduction}

The common envelope (CE) phase was proposed by \citet{paczynski76} to
explain the small orbital separation of compact binaries such as
cataclysmic variables (CVs) and double white dwarf binaries, because a
much larger orbital separation of between approximately 10 and 1000
R$_{\odot}$ was required in order to accommodate the giant progenitor
of the white dwarf. (For reviews see Iben \& Livio 1993, Taam \&
Sandquist 2000).

In the pre-CE phase of evolution the initially more massive stellar
component (which we henceforth denote as the primary) evolves off the
main sequence first. Depending on the orbital separation of the
binary, the primary will fill its Roche lobe on either the giant or
asymptotic giant branch and initiate mass transfer. If the giant
primary possesses a deep convective envelope (i.e. the convective
envelope has a mass of more than approximately 50 per cent of the
giant's mass; Hjellming \& Webbink 1987) the giant will expand in
response to rapid mass loss. As a result, the giant's radius expands
relative to its Roche lobe radius increasing the mass transfer
rate. As a consequence of this run-away situation, mass transfer
commences on a dynamical timescale. The companion main sequence star
(henceforth the secondary) 
cannot incorporate this material into
its structure 
quickly enough
and therefore expands to fill its own Roche lobe. The envelope
eventually engulfs both the core of the primary and the main sequence
secondary.

The friction the stellar components experience due to their motion
within the CE extracts energy and orbital angular momentum from the
orbit causing the orbital separation to decrease. If enough of the
orbital energy is imparted on to the CE before the stellar components
merge, then the CE may be ejected from the system leaving the core of
the primary (now the white dwarf) and the secondary star at a greatly
reduced separation.

Despite extensive three dimensional hydrodynamical simulations
(e.g. Sandquist, Taam \& Burkert 2000), the physics of the CE phase
remains poorly understood, including the efficiency with which the CE
is ejected from the system. Due to the difficulty in modelling the CE
phase, binary population synthesis calculations 
commonly resort to describing the CE phase in terms of a simple energy budget
argument. A fraction $\alpha_{\rmn{CE}}$ of the orbital energy released as the binary tightens, 
$\Delta{E_{\rmn{orb}}}$, is available to unbind the giant's envelope from the core. 
Hence if the binding energy is $E_{\rmn{bind}}$ we have
\begin{equation}
E_{\rmn{bind}} = \alpha_{\rmn{CE}} \Delta{E}_{\rmn{orb}}
\label{Eorb}
\end{equation}
(e.g. de Kool 1992; de Kool \& Ritter 1993; Willems \& Kolb 2004).
The efficiency $\alpha_{\rmn{CE}}$ is a free parameter with values $0<\alpha_{\rmn{CE}}\le{1}$, albeit 
values larger than unity are discussed, depending on what terms are included in the quantity 
$E_{\rmn{bind}}$.

\citet{nelemans00} 
used this approach to reconstruct 
the common envelope phase for the
observed sample of double white dwarf binaries. They found that
$\alpha_{\rmn{CE}}<0$ for the first phase of mass transfer, which is
unphysical. This led them to test an alternative description of 
the CE phase, based on the
change in the orbital angular momentum of the binary, $\Delta{J}$,
during the CE phase. Their `$\gamma$-algorithm' follows
the prescription given by \citet{pz67},
\begin{equation}
\Delta{J}=\gamma\frac{JM_{\rmn{env}}}{M_{\rmn{b}}},
\label{nelemans1}
\end{equation}
where $M_{\rmn{env}}$ is the mass of the giant's envelope,
$M_{\rmn{b}}$ is the total mass of the binary just before the onset of
the CE phase and $J$ is the total angular momentum of the binary at
this point. The free parameter $\gamma$ describes the specific angular
momentum of the ejected material (in this case the CE) in units of the
specific angular momentum of the binary system. Indeed, \citet{nt05}
found further support for the $\gamma$-algorithm in a few
observed pre-CVs and sub-dwarf B plus main sequence star
binaries.

\citet{beer07} suggested that some systems may avoid the 
in-spiral in a CE phase
altogether as a result of wind ejection of the envelope material from the system,
which arises due to super-Eddington accretion on to the companion
star. Employing analytical approximations for the evolution of such systems, 
\citet{beer07} found that both the canonical energy budget approach and the
$\gamma$-algorithm can describe the outcomes of the proposed mechanism. 
%
%
%

A few observed double white dwarf binaries could also be explained by
$\alpha_{\rmn{CE}}>1$, for example, PG 1115+166 \citep{maxted02}. This
has led to the suggestion that an energy source other than the release
of orbital energy is being used to eject the CE from the system, such
as thermal energy and recombination energy of ionized material within
the giant's envelope (Han et al. 1994, Han et al. 1995, Dewi \& Tauris
2000, Webbink 2007).

Alternatively, \citet{pw07} suggested that rather than
$\alpha_{\rmn{CE}}$ being a constant, global value for all binary
systems, it is a function of one of the binary orbital
parameters. 
In this spirit, \citet{pw07} calculated the present day population of
post-common envelope binaries (PCEBs) and CVs if 
$\alpha_{\rmn{CE}}$ is a function of the secondary mass. The
investigation was motivated by the fact that very few, if any, CVs  
with brown dwarf secondaries at orbital periods below 77 mins have been
detected.
This 
appears to be in conflict with \citet{politano04} who estimated from 
his models 
that such systems should make up approximately 15 per cent of the
current CV population. 
\citet{politano04} suggested that this
discrepancy may be a result of the decreasing energy dissipation rate
of orbital energy within the CE for decreasing secondary mass, and
that below some cut-off mass, a CE merger would be unavoidable.

Previous population synthesis studies into the CE phase of white
dwarf-main sequence binaries(e.g. de Kool \& Ritter 1993; Willems \&
Kolb 2004) just considered the value of the envelope's binding energy,
$E_{\rmn{bind}}$, due to the gravitational binding energy of the
primary's envelope,
\begin{equation}
E_{\rmn{bind}}=-\int^{M_{1}}_{M_{\rmn{c}}}\frac{GM(r)}{r}\,\rmn{d}m,
\label{Ebind}
\end{equation}
where $M(r)$ is the mass contained within the radius $r$ of the
primary, $M_{\rmn{c}}$ is the mass of the giant's core, and $G$ is the
universal gravitational constant. In population synthesis studies 
the binding energy is commonly approximated as 
\begin{equation}
E_{\rmn{bind}} = - \frac{GM_{1}M_{\rmn{env}}}{\lambda R_{1}},
\label{Ebind0}
\end{equation}
where $R_{1}$ is the radius of the primary star, with a suitable choice for
the dimensionless factor $\lambda$.

\citet{deKool92} and \citet{wk04} for example, used a constant, global
value of $\lambda=0.5$. As \citet{dt00} pointed out, however, this may
lead to an overestimation of the giant envelope's binding energy for
large radii, requiring more energy to eject it.

In the present study we apply population synthesis techniques to calculate 
the present
day population of PCEBs by considering the aforementioned theoretical
descriptions of the CE phase. We consider different constant, global
values of $\alpha_{\rmn{CE}}$ and $\lambda$, $\alpha_{\rmn{CE}}$ as a
function of secondary mass, and $\lambda$ calculated according to the
internal structure of the giant star and the internal energy of its
envelope. Also, we consider the description of the CE phase is terms
of the angular momentum budget of the binary system.

We then compare our model PCEB populations to the observed sample of
white dwarf-main sequence star (WD+MS) and sub-dwarf-main sequence
star (sd+MS) binaries from RKCat \citep{rk03}, Edition 7.10 (2008),
and newly detected PCEBs from the Sloan Digital Sky Survey (SDSS)
(Rebassa-Mansergas et al. 2007; Rebassa-Mansergas et al. 2008). We
also compare this observational sample to our theoretical population
for a range of initial secondary mass distributions.

Our suite of population models represents a major advance over the
work by \citet{wk04} who considered only the formation rate of PCEBs,
and a simplified estimate of their present-day populations based on
birth and death rates. Our models also cover a more comprehensive
parameter space than those by \citet{pw07}, and we assess the model
distributions against the location of observed systems in all three
system parameters (component masses and period) simultaneously.

The structure of the paper is as follows. In Section 2 we give our
computational method and models describing the CE phase. In Section 3
we present our results, which are discussed in Section 4. We conclude
our investigation in Section 5.

\section{Computational Method}

We employ the same method as used in \citet{dkwg08} to calculate the
binary population, and so we present only a summary here. We first
calculate the unweighted zero-age PCEB population (i.e. WD+MS systems
that have just emerged from a CE phase) with BiSEPS (Willems \& Kolb
2002; Willems \& Kolb 2004). BiSEPS employs the single star evolution
(SSE) formulae described in \citet{hpt00} and a binary evolution
scheme based on that described by \citet{htp02}. We then use a second
code introduced in \citet{willems05} to calculate the present day
population of PCEBs.

This latter step makes use of evolutionary tracks, calculated by BiSEPS,  
for each PCEB configuration, from the moment the PCEB forms to
the point at which the binary ceases to be a PCEB (for
example, if the system becomes semi-detached). 
BiSEPS therefore produces a library of PCEB evolutionary tracks which contain, 
depending on the population model, between $30\,000$ and $100\,000$ PCEB sequences.

\subsection{Initial Binary Population}

BiSEPS evolves a large number of binary systems, initially consisting
of two zero-age main sequence stellar components. The stars are
assumed to have a population I chemical composition and the orbits are
circular at all times. The initial primary and secondary masses are in
the range 0.1 to 20 M$_{\odot}$, while the initial orbital periods
range from 0.1 to 100 000 d. There is one representative binary
configuration per grid cell within a three-dimensional grid consisting
of 60 logarithmically spaced points in primary and secondary mass and
300 logarithmically spaced points in orbital period. Hence we evolve
approximately $5.4\times{10}^{5}$ binaries for a maximum evolution
time of 10 Gyr. For symmetry reasons only systems where $M_{1}>M_{2}$
are evolved.

The probability of a zero-age PCEB forming with a given white dwarf
mass $M_{\rmn{WD}}$, secondary mass $M_{2}$ and orbital period
$P_{\rmn{orb}}$ is determined by the probability of the binary's
initial parameters. We assume that the initial primary mass,
$M_{1,\rmn{i}}$ is distributed according to the initial mass function
(IMF) given by \citet{ktg93}. The distribution of initial secondary
masses, $M_{2,\rmn{i}}$, is obtained from the initial mass ratio
distribution (IMRD), $n(q_{\rmn{i}})$, where
$q_{\rmn{i}}=M_{2,\rmn{i}}/M_{1,\rmn{i}}$, and $n(q_{\rmn{i}})$ has a
power-law dependence on $q_{\rmn{i}}$. Alternatively, we determine
$M_{2,\rmn{i}}$ from the same IMF as used for the initial primary
mass. For brevity, we adopt the acronym `IMFM2' for this latter
case. Our standard model is a flat distribution,
i.e. $n(q_{\rmn{i}})=1$. Finally, we follow \citet{it84} and
\citet{htp02} and adopt a logarithmically flat distribution in initial
orbital separations.

We assume that all stars in the Galaxy are formed in binaries, and
that the Galaxy has an age of 10 Gyr. The aforementioned distributions
are then convolved with a constant star formation rate normalised such
that one binary with $M_{1,\rmn{i}}>0.8$ M$_{\odot}$ is formed each
year, consistent with observations \citep{weidemann90}. This gives an
overall star formation rate of $7.6$ yr$^{-1}$.

\begin{table}
  \centering
    \caption{Different models on the treatment of the CE phase.}
    \begin{tabular}{@{}lll@{}}
    \hline
    Model   &   $\alpha_{\rmn{CE}}$ or $\gamma$                                         &  $\lambda$  \\
    \hline
    A       &   1.0                                                                     &  0.5        \\
    CE01    &   0.1                                                                     &  0.5        \\
    CE06    &   0.6                                                                     &  0.5        \\
    \hline
    PL05    &   eqn. (\ref{alpha_func_m2}), $p=0.5$,                                    &  0.5       \\
    PL1     &   eqn. (\ref{alpha_func_m2}), $p=1$,                                      &  0.5      \\
    PL2     &   eqn. (\ref{alpha_func_m2}), $p=2$,                                      &  0.5      \\
    \hline
    CT0375  &   eqn. (\ref{alpha_func_mcut}), $M_{\rmn{cut}}/\rmn{M}_{\odot}=0.0375$    &  0.5 \\
    CT075   &   eqn. (\ref{alpha_func_mcut}), $M_{\rmn{cut}}/\rmn{M}_{\odot}=0.075$     &  0.5 \\
    CT15    &   eqn. (\ref{alpha_func_mcut}), $M_{\rmn{cut}}/\rmn{M}_{\odot}=0.15$      &  0.5 \\
    \hline
    DTg     &   1.0                                             &  $\lambda_{\rmn{g}}$  \\
    DTb     &   1.0                                             &  $\lambda_{\rmn{b}}$  \\
    \hline
    n15     &   1.5, eqn. (\ref{nelemans2})                          &  -                                  \\
\hline
\end{tabular}
\label{models}
\end{table}

\subsection{The Common Envelope Phase}

Here we specify details of the different model descriptions used for
the CE phase. These are also summarised in Table \ref{models}. In the
energy budget approach the CE phase is modelled by equating the change
in the binding energy of the primary's envelope to the change in the
orbital energy of the binary system (see Section 1,
eqn. \ref{Eorb}). Thus, the final orbital separation of the binary
after the CE phase, $A_{\rmn{CE,f}}$, is given by
\begin{equation}
A_{\rmn{CE,f}}=A_{\rmn{CE,i}}\frac{M_{\rmn{c}}}{M_{1}}\left[1+\left(\frac{2}{\lambda\alpha_{\rmn{CE}}r_{\rmn{L},1}}\right)\left(\frac{M_{\rmn{env}}}{M_{2}}\right)\right]^{-1}.
\label{af}
\end{equation}
Here $A_{\rmn{CE,i}}$ is the initial orbital separation at the onset
of the CE phase, $M_{1}$ and $M_{2}$ are the primary and secondary
masses just at the start of the CE phase respectively, $M_{\rmn{c}}$
is the primary's core mass,
$r_{\rmn{L},1}\equiv{R_{\rmn{L},1}}/A_{\rmn{CE,i}}$ is the radius of
the primary star in units of the orbital separation at the start of
the CE phase and $M_{\rmn{env}}$ is the mass of the giant's
envelope. For our standard model we use $\lambda=0.5$, which is
consistent with \citet{wk04}. This forms our model A.

We consider three constant, global values of $\alpha_{\rmn{CE}}$. For
our reference model A we use $\alpha_{\rmn{CE}}=1$. We also consider
$\alpha_{\rmn{CE}}=0.1$ and 0.6. These models are denoted as CE01 and
CE06 respectively, where the last two digits correspond to the value
of $\alpha_{\rmn{CE}}$.

Following \citet{pw07} we consider $\alpha_{\rmn{CE}}$ as a function
of the secondary mass. We first adopt
\begin{equation}
\alpha_{\rmn{CE}}=\left(\frac{M_{2}}{\rmn{M}_{\odot}}\right)^p,
\label{alpha_func_m2}
\end{equation}
where $p$ is a free parameter. We consider $p=0.5$, 1.0 and 2.0. These
models are denoted as PL05, PL1 and PL2 respectively, where the last
digits correspond to the value of $p$.

For completeness, we also consider the second functional form given by
\begin{equation}
  \alpha_{\rmn{CE}} = \left\{
  \begin{array}{l l}

  1-\frac{M_{\rmn{cut}}}{M_{2}} & \quad \mbox{$M_{2}>M_{\rmn{cut}}$}\\
  0 & \quad \mbox{$M_{2}\le{M_{\rmn{cut}}}$}\\

  \end{array} \right.
\label{alpha_func_mcut}
\end{equation}
where $M_{\rmn{cut}}$ is a cut-off mass. As \citet{pw07} we use
$M_{\rmn{cut}}=0.0375$, 0.075 and 0.15, which corresponds to
$\times{1/2}$, $\times{1}$ and $\times{2}$ of the sub-stellar mass
respectively. These models are denoted as CT0375, CT075 and CT15,
where the digits correspond to the cut-off mass.

We also obtain population models with values of $\lambda$ calculated
by \citet{dt00}, where just the gravitational binding energy of the
envelope is considered (their $\lambda_{\rmn{g}}$) and where both the
gravitational binding energy and the thermal energy of the envelope
are considered (their $\lambda_{\rmn{b}}$). We denote these models as
DTg and DTb respectively (and set $\alpha_{\rmn{CE}}=1$ for these).

For each of our binary configurations we calculate the value of
$\lambda_{\rmn{g}}$ or $\lambda_{\rmn{b}}$ by applying linear
interpolation of the values for $M_{1}$, $R_{1}$ and
$\lambda_{\rmn{g,b}}$ tabulated by \citet{dt00}. However, \citet{dt00}
only tabulate $\lambda$ values 
for primary masses $M_{1}\ge{3}$ M$_{\odot}$, and for primary radii up
to between 400 and 600 R$_{\odot}$. 
We extended this table using full stellar models
calculated by the Eggleton code (provided by van der Sluys, priv.
comm.), up to the point where the dynamical timescale of the star becomes
too small, and the evolutionary calculations stop. 

\begin{figure}
  \centering
  \includegraphics[scale=0.55]{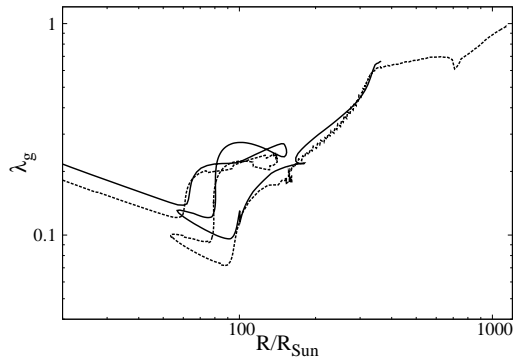}
  \caption{Values of $\lambda_{\rmn{g}}$ with stellar radius as
  calculated by the Eggleton code (solid line) and by the EZ code
  (dashed line), for a 6 M$_{\odot}$ star.}
  \label{lambda_R}
\end{figure}

In particular, for the 1 M$_{\odot}$ stellar model, we obtain values of
$\lambda_{\rmn{g}}$ and $\lambda_{\rmn{b}}$ for radii up to
approximately 500 R$_{\odot}$. For stellar masses between 
2 and 4 M$_{\odot}$ we obtain $\lambda$ values up to approximately
1000 R$_{\odot}$, while for stellar masses between 5 and
8 M$_{\odot}$\footnote{The upper limit of 8 M$_{\odot}$ is the largest primary
progenitor which will form a white dwarf.} up to
between approximately 400 and 600 R$_{\odot}$.




However, we find that PCEB progenitors with primary masses
between 2 and 8 M$_{\odot}$ can fill their Roche lobes with radii up
to about 1200 R$_{\odot}$. To estimate likely $\lambda$ values for
these large radii we used the EZ stellar evolution code, which is a
re-written, stripped down adaptation of the Eggleton code
\citep{paxton04}. Indeed, the code was designed to be more stable and
faster, though at the cost of some detail. In the mass range of
approximately 5 to 6 M$_{\odot}$ this does return $\lambda_{g}$ values
for large radii\footnote{For these masses, $\lambda_{\rmn{b}}<0$ at
radii $R_{1}\ga{500}.$} (see Fig. \ref{lambda_R} for the 6 M$_{\odot}$
star).  However, for masses below 5 M$_{\odot}$ or above 6
M$_{\odot}$, the EZ code stops the evolution at radii similar to the
Eggleton code. For these stellar masses, we simply take the last
calculated values of $\lambda_{\rmn{g}}$ for any $R_{1}>600$
R$_{\odot}$ in our population models.

If there is sufficient thermal energy within the giant's envelope such
that its total binding energy becomes positive ($\lambda_{\rmn{b}}<0$)
the formalism described by equation (\ref{af}) breaks down. We model
this instead as an instantaneous ejection of the giant's envelope via
a wind, which takes away the specific orbital angular momentum of the
giant. What remains is a wide WD+MS binary. We do not consider these
any further in this investigation.

\subsection{CE Phase in terms of Angular Momentum}

We follow \citet{nelemans00} and \citet{nt05} by considering models
where the CE phase is described in terms of the change of the binary's
angular momentum, $\Delta{J}$, given by equation (\ref{nelemans1}) in
Section 1. If $J$ and $J_{\rmn{f}}$ are the total angular momentum of
the binary immediately before and after the CE phase respectively,
then equation (\ref{nelemans1}) can be re-written as
\begin{equation}
J\left[1-\gamma\left(\frac{M_{1}-M_{\rmn{c}}}{M_{1}+M_{2}}\right)\right]=J_{\rmn{f}}.
\label{nelemans2}
\end{equation}
At the start of the CE phase, we not only consider the orbital angular
momentum of the binary, $J_{\rmn{orb,i}}$, but also the spin angular
momentum of the Roche lobe-filling giant star, $J_{\rmn{spin,1}}$,
such that $J=J_{\rmn{spin,1}}+J_{\rmn{orb,i}}$. If the giant is
rotating synchronously with the orbit with a spin angular speed of
$\Omega_{\rmn{orb}}$, then
\begin{equation}
J_{\rmn{spin,1}}=k^{2}M_{1}R_{\rmn{L},1}^{2}\Omega_{\rmn{orb}},
\label{Jspin}
\end{equation}
where $k^{2}$ is the ratio of the gyration radius to the radius
of the giant star. We use the values of $k^{2}$ tabulated by
\citet{dt00}. Our procedure for calculating $k^{2}$ for each giant
star configuration is the same as that used to calculate
$\lambda_{\rmn{g}}$ and $\lambda_{\rmn{b}}$. The initial orbital
angular momentum of the binary, on the other hand is given by
\begin{equation}
J_{\rmn{orb,i}}=M_{1}M_{2}\left(\frac{GA_{\rmn{CE,i}}}{M_{1}+M_{2}}\right)^{1/2}.
\label{Jorbi}
\end{equation}
Similarly the final orbital angular momentum of the binary after the
CE phase, $J_{\rmn{orb,f}}$, is given by
\begin{equation}
J_{\rmn{orb,f}}=M_{\rmn{c}}M_{2}\left(\frac{GA_{\rmn{CE,f}}}{M_{\rmn{c}}+M_{2}}\right)^{1/2}.
\label{Jorbf}
\end{equation}
Note that we do not consider the spin angular momentum of either of
the stellar components after the CE phase as these are negligible
compared to the orbital angular momentum of the binary. Thus,
$J_{\rmn{f}}=J_{\rmn{orb,f}}$. From equations (\ref{Jspin}) to
(\ref{Jorbf}) we can therefore solve for $A_{\rmn{CE,f}}$ in equation
(\ref{nelemans2}).

For the large majority of the observed systems that \citet{nt05}
considered, they found that their CE phase could be reproduced with
$\gamma\approx{1.5}$. We therefore consider $\gamma=1.5$ in equation
(\ref{nelemans2}). This is our model n15, which is summarised in Table
\ref{models}.

\subsection{Magnetic Braking}

After the CE phase, the subsequent evolution of the PCEB will be
driven by angular momentum losses from the orbit. We follow the
canonical disrupted magnetic braking 
hypothesis, 
and assume that for
a fully convective secondary star with
$M_{2}\le{M_{\rmn{conv,MS}}=0.35}$ M$_{\odot}$ (where
$M_{\rmn{conv,MS}}$ is the maximum mass of a fully convective,
isolated, 
zero-age 
main sequence star), gravitational radiation is the only
sink of angular momentum. For secondaries with
$M_{2}>M_{\rmn{conv,MS}}$ on the other hand, we assume that the
evolution of the PCEB is driven by a combination of gravitational
radiation and magnetic braking. For this investigation we consider
the angular momentum loss rate due to magnetic braking,
$\dot{J}_{\rmn{MB}}$, as given by \citet{htp02}
\begin{equation}
\dot{J}_{\rmn{MB}}=-\eta_{\rmn{h}}5.83\times{10}^{-16}\rmn{M_{\odot}R_{\odot}^{-1}yr}\frac{M_{\rmn{conv}}}{M_{2}}R_{2}^{3}\Omega^{3},
\label{MB_hurley}
\end{equation}
where $R_{2}$ is the secondary's radius, $M_{\rmn{conv}}$ is the mass
of the secondary's convective envelope and $\Omega$ is its spin
frequency. We normalise equation (\ref{MB_hurley}) as described in
\citet{dkwg08} by applying a factor $\eta_{\rmn{h}}=0.19$ which gives
the angular momentum loss rate appropriate for a period gap of the
observed width of one hour. Magnetic braking also becomes ineffective
for PCEBs with $M_{2}\ge{1.25}$ M$_{\odot}$, which have very thin or
no convective envelopes.

\section{Results}

We begin by comparing our theoretical population of PCEBs from model
A, $n(q_{\rmn{i}})=1$, with observed WD+MS and sd+MS systems, obtained
from Edition 7.10 (2008) of RKCat \citep{rk03}. This sample is also
supplemented with newly discovered PCEBs from SDSS (Rebassa-Mansergas
et al. 2007; Rebassa-Mansergas et al. 2008). Observed WD+MS systems
are tabulated in Table \ref{table01}, while observed sd+MS systems are
tabulated in Table \ref{table02}. We then discuss the present day
number and local space densities of PCEBs, and discuss each of the CE
models in turn.

\begin{table*}
  \centering
  \begin{minipage}{170mm}

    \caption{The orbital periods, white dwarf masses ($M_{\rmn{WD}}$),
    and secondary masses, $M_{2}$, of the observed sample of WD+MS
    systems. Also shown are the weightings of each system in the
    associated white dwarf mass range, corresponding to panels (a) to
    (i) (see Fig. \ref{multiplot}).}

    \begin{tabular}{@{}llllllll@{}}
    \hline
    Name        &  Alt. name   &   $P_{\rmn{orb}}$/ d   &   $M_{\rmn{WD}}$   &   $M_{2}$       &   Panel      &    Weighting (\%)            & Ref. \\
                &              &                        &    /M$_{\odot}$    & /M$_{\odot}$    &              &                              & \\
    \hline
    V651 Mon    &  AGK3-0 965  &     15.991             &   0.40$\pm{0.05}$  & 1.80$\pm{0.3}$   &  a,b         & 50.0,48.0                 &  1,2  \\
    SDSS J1529+0020 &          &     0.165              &   0.40$\pm{0.04}$  & 0.25$\pm{0.12}$  &  a,b         & 50.0,49.0                 &  3    \\
    CC Cet      &  PG 0308+096 &     0.287              &   0.40$\pm{0.11}$  & 0.18$\pm{0.05}$  &  a,b,c       & 50.0,32.0,1.00            &  4   \\
    LM Com      & CBS 60       &     0.259              &   0.35$\pm{0.03}$  & 0.17$\pm{0.02}$  &  a           & 95.0                      &  5\\
    HR Cam      & WD 0710+741  &     0.103              &   0.41$\pm{0.01}$  & 0.096$\pm{0.004}$&  a,b         & 16.0,84.0                 &  6,7\\
    0137-3457   &              &     0.080              &   0.39$\pm{0.035}$ & 0.053$\pm{0.006}$&  a,b         & 61.0,39.0                 &  8\\
    SDSS J2339-0020 &          &     0.656              &   0.8$\pm{0.4}$    & 0.32$\pm{0.09}$  &  a,b,c,d     & 14.0,7.0,4.0,4.0          &  3\\
                    &          &                        &                    &                  &  e,f,g,h,i   & 5.0,5.0,10.0,27.0,17.0    &  \\
    SDSS J1724+5620 &          &     0.333              &   0.42$\pm{0.01}$  & 0.25 - 0.38      &  b           & 98.0                      &  3\\
    RR Cae      & LFT 349      &     0.303              &   0.44$\pm{0.023}$ & 0.18$\pm{0.01}$  &  b           & 95.0                      &  9\\
    MS Peg      & GD 245       &     0.174              &   0.49$\pm{0.04}$  & 0.19$\pm{0.02}$  &  b,c         & 59.0,33.0                 &  10\\
    HZ 9        &              &     0.564              &   0.51$\pm{0.1}$   & 0.28$\pm{0.04}$  &  b,c,d,e     & 32.0,10.0,16.0,10.0       &  11,12,13\\
    GK Vir      & PG 1413+015  &     0.344              &  0.51$\pm{0.04}$   & 0.10             &  b,c         & 40.0,44.0                 &  14,15\\
    LTT 560     &              &     0.148              &  0.52$\pm{0.12}$   & 0.19$\pm{0.05}$  &  b,c,d,e     & 28.0,16.0,15.0,11.0       &  16\\
    DE CVn      & J1326+4532   &     0.364              &  0.54$\pm{0.04}$   & 0.41$\pm{0.06}$  &  c,d         & 44.0,33.0                 &  17\\
    2237+8154   &              &     0.124              &  0.57$\pm{0.1}$    & 0.30$\pm{0.1}$   &  b,c,d,e,f   & 20.0,18.0,20.0,17.0,12.0  &  18\\
    SDSS J1151-0007 &          &     0.142              &   0.6$\pm{0.1}$    & 0.19$\pm{0.08}$  &  b,c,d,e,f   & 14.0,15.0,19.0,19.0,15.0  &  3 \\
    UZ Sex      & 1026+0014    &     0.597              &  0.68$\pm{0.23}$   & 0.22$\pm{0.05}$  &  b,c,d,e,    & 11.0,6.90,7.80,8.40,      &  4\\
                &              &                        &                    &                  &  f,g,h       & 8.7,16.0,27.0             &   \\
    NN Ser      & PG 1550+131  &    0.130               &  0.54$\pm{0.04}$   & 0.150$\pm{0.008}$&  c,d         & 44.0,33.0                 &  19\\
    FS Cet      & Feige 24     &    4.232               &  0.57$\pm{0.03}$   & 0.39$\pm{0.02}$  &  c,d         & 24.0,59.0                 &  20,21\\
    J2013+4002  &              &    0.706               &  0.56$\pm{0.03}$   & 0.23$\pm{0.01}$  &  c,d         & 35.0,54.0                 &  21,22\\
    J2130+4710  &              &    0.521               &  0.554$\pm{0.017}$ & 0.555$\pm{0.023}$&  c,d         & 41.0,59.0                 &  23\\
    1042-6902   &              &    0.337               &  0.56$\pm{0.05}$   & 0.14$\pm{0.01}$  &  c,d,e       & 31.0,37.0,18.0            &  21\\
    SDSS J0314-0111 &          &    0.263               &  0.65$\pm{0.1}$    & 0.32$\pm{0.09}$  &  c,d,e,f,g   & 9.0,15.0,19.0,19.0,24.0   &  3 \\
    IN CMa      & J0720-3146   &    1.262               &  0.58$\pm{0.03}$   & 0.43$\pm{0.03}$  &  d,e         & 59.0,24.0                 &  24\\
    J1016-0520AB&              &    0.789               &  0.61$\pm{0.06}$   & 0.15$\pm{0.02}$  &  d,e,f       & 28.0,31.0,19.0            &  24\\
    EG UMa      & Case 1       &    0.668               &  0.63$\pm{0.05}$   & 0.36$\pm{0.04}$  &  d,e,f       & 22.0,38.0,26.0            &  25\\
    2009+6216   &              &    0.741               &  0.62$\pm{0.02}$   & 0.189$\pm{0.004}$&  e           & 77.0                      &  26\\
    BE UMa      & PG 1155+492  &    2.291               &  0.70$\pm{0.07}$   & 0.36$\pm{0.07}$  &  e,f,g       & 16.0,26.0,42.0            &  27,28\\
    1857+5144   &              &    0.266               &  0.80$\pm{0.2}$    & 0.23$\pm{0.08}$  &  e,f,g,h     & 6.80,8.20,19.0,43.0       &  29\\
    SDSS J0246+0041 &          &    0.726               &  0.9$\pm{0.2}$     & 0.38$\pm{0.07}$  &  f,g,h       & 5.0, 15.0,53.0            &  3\\
    SDSS J0052-0053 &          &    0.114               &  1.2$\pm{0.4}$     & 0.32$\pm{0.09}$  &  g,h,i       & 5.0,24.0,32.0             &  3\\
    BPM 71214   &              &    0.202               &  0.77$\pm{0.06}$   & 0.540:           &  g,h         & 57.0,31.0                 &  30\\
    QS Vir      & 1347-1258    &    0.151               &  0.78$\pm{0.04}$   & 0.43$\pm{0.04}$  &  g,h         & 67.0,31.0                 &  31\\
    V471 Tau    & BD+16 516    &    0.521               &  0.84$\pm{0.05}$   & 0.93$\pm{0.07}$  &  g,h         & 21.0,79.0                 &  32\\
    IK Peg      & BD+18 4794   &    21.72               &  1.19$\pm{0.05}$   & 1.7              &  i           & 96.0                      &  33\\

    \hline
\label{table01}
\end{tabular}
References: (1)\,\citet{mn81}, (2)\,\citet{mendez85}, (3)\,\citet{rebassa08}, (4)\,\citet{saffer93}, (5)\,\citet{shimansky03}, (6)\,\citet{md96},
 (7)\,\citet{maxted98}, (8)\,\citet{maxted06}, (9)\,\citet{bragaglia95}, (10)\,\citet{schmidt95}, (11)\,\citet{stauffer87}, (12)\,\citet{lp81},
 (13)\,\citet{gs84}, (14)\,\citet{fulbright93}, (15)\,\citet{green78}, (16)\,\citet{tappert07}, (17)\,\citet{besselaar07}, (18)\,\citet{gaensicke04},
 (19)\,\citet{catalan94}, (20)\,\citet{vt94}, (21)\,\citet{kawka08}, (22)\,\citet{good05}, (23)\,\citet{maxted04},
 (24)\,\citet{vennes99}, (25)\,\citet{bleach00}, (26)\,\citet{morales05}, (27)\,\citet{wood95}, (28)\,\citet{ferguson99}, (29)\,\citet{aungwerojwit07},
 (30)\,\citet{kawka02}, (31)\,\citet{odonoghue03}, (32)\,\citet{obrien01}, (33)\,\citet{lsb93}
\end{minipage}
\end{table*}

\begin{table*}
  \centering
  \begin{minipage}{170mm}
    \caption{The orbital periods, sub-dwarf mass ($M_{\rmn{SD}}$),
    secondary mass, $M_{2}$, and spectral type, SpT, of the sub-dwarf
    for the observed sample of sd+MS systems. Also shown are the
    weightings of each system in the corresponding panels (a) to (f)
    (see Fig. \ref{multiplot2}), except for those systems where no
    uncertainty in the sub-dwarf mass is available. A colon denotes an
    uncertainty flag.}
    \begin{tabular}{@{}lllllllll@{}}
    \hline
    Name          &  Alt. name   &   $P_{\rmn{orb}}$/ d   &   $M_{\rmn{SD}} $  & SpT 1  &  $M_{2}$          &  Panel  &    Weighting (\%)    & Ref. \\
                  &              &                        &    /M$_{\odot}$    &        &  /M$_{\odot}$     &         &                      &      \\
    \hline
    V1379 Aql     & HD 185510    &   20.662               &  0.304$\pm{0.015}$ & sdB   & 2.27$\pm{0.13}$   &  a       & 100.0                & 1 \\
    FF Aqr        & BD -3 5357   &   9.208                &  0.35$\pm{0.06}$ : & sdOB  & 1.40$\pm{0.4}$ :  &  a,b     & 80.0,20.0            & 2,3 \\
    2333+3927     &              &   0.172                &  0.38$\pm{0.09}$   & sdB   & 0.28$\pm{0.04}$   &  a,b     & 59.0,32.0            & 4 \\
    AA Dor        & LB 3459      &   0.262                &  0.330$\pm{0.006}$ & sdO   & 0.066$\pm{0.001}$ &  a       & 100.0                & 5 \\
    HW Vir        & BD -7 3477   &   0.117                &  0.48$\pm{0.09}$   & sdB   & 0.14$\pm{0.02}$   &  a,b,c,d & 19.0,40.0,19.0,13.0  & 6 \\
    2231+2441     &              &   0.111                &  0.47 :            & sdB   & 0.075             &  b       & -                    & 7 \\
    NY Vir        & PG 1336-018  &   0.101                &  0.466$\pm{0.006}$ & sdB   & 0.122$\pm{0.001}$ &  b       & 100.0                & 8 \\
    0705+6700     &              &   0.096                &  0.483             & sdB   & 0.134             &  b       & -                    & 9,10\\
    BUL-SC 16 335 &              &   0.125                &  0.5 :             & sdB   & 0.16              &  b       & -                    & 11\\
    J2020+0437    &              &   0.110                &  0.46              & sdB   & 0.21              &  b       & -                    & 12\\
    XY Sex        & 1017-0838    &   0.073                &  0.50              & sdB   & 0.078$\pm{0.006}$ &  b       & -                    & 13\\
    V664 Cas      & HFG 1        &   0.582                &  0.57$\pm{0.03}$   & sdB   & 1.09$\pm{0.07}$   &  c,d     & 24.0,59.0            & 14,15,16\\
    UU Sge        &              &   0.465                &  0.63$\pm{0.06}$   & sdO   & 0.29$\pm{0.04}$   &  d,e,f   & 22.0,32.0,25.0       & 17\\
    KV Vel        & LSS 2018     &   0.357                &  0.63$\pm{0.03}$   & sdO   & 0.23$\pm{0.01}$   &  d,e,f   & 15.0,59.0,24.0       & 18\\
    \hline
\label{table02}
\end{tabular}

References: (1)\,\citet{js97}, (2)\,\citet{vw03},
(3)\,\citet{etzel77}, (4)\,\citet{heber04}, (5)\,\citet{rauch00},
(6)\,\citet{ws99}, (7)\,\citet{ostensen07}, (8)\,\citet{kilkenny98},
(9)\,\citet{drechsel01}, (10)\,\citet{nemeth05},
(11)\,\citet{polubek07}, (12)\,\citet{wils07}, (13)\,\citet{maxted02},
(14)\,\citet{miller76}, (15)\,\citet{pb93}, (16)\,\citet{bell94},
(17)\,\citet{hilditch96}, (18)\,\citet{shimanskii04}

\end{minipage}
\end{table*}

\subsection{White dwarf-main sequence systems}

\begin{figure*}
  \centering
  \begin{minipage}{175mm}
    \includegraphics[scale=0.9]{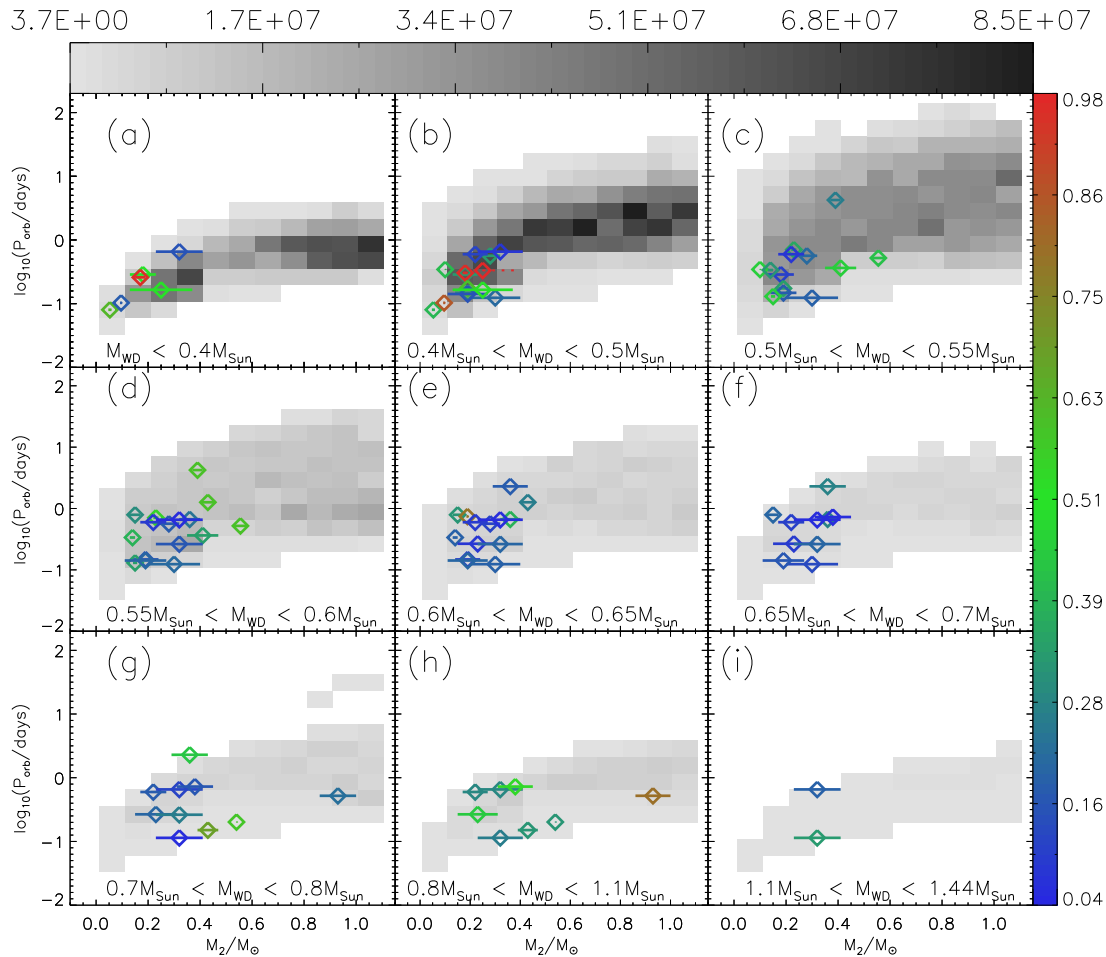}
    \caption{Calculated present-day PCEB populations, for model A and
    for $n(q_{\rmn{i}})=1$. The distributions over $M_{2}$ and
    $\rmn{log}P_{\rmn{orb}}$ are shown for nine different white dwarf
    mass intervals as indicated in each of the nine panels, labelled
    (a) to (i). The grey-scale bar at the top of the plot indicates
    the number of systems per bin area. Also shown are the known WD+MS
    systems obtained from Edition 7.10 (2008) of RKCat
    \citep{rk03}. We take into account the uncertainties of the white
    dwarf mass for each system by calculating the weighting of each
    system within each panel. This is repeated for those systems whose
    error bars in white dwarf mass overlaps more than one panel. The
    weighting is indicated by the colour bar on the right hand side of
    the plot. The dotted line in panel (b) indicates that the
    secondary mass in SDSS J1724+5620 lies between 0.25 M$_{\odot}$
    and 0.38 M$_{\odot}$.}
    \label{multiplot}
  \end{minipage}
\end{figure*}

Figure \ref{multiplot} shows the theoretical present-day PCEB
populations on the $M_{2}-\rmn{log}\,P_{\rm{orb}}$ plane for model A,
assuming an IMRD of $n(q_{\rmn{i}})=1$, along with the observed WD+MS
systems.

\begin{figure*}
  \centering
  \begin{minipage}{175mm}
    \includegraphics*[viewport=0 240 500 420]{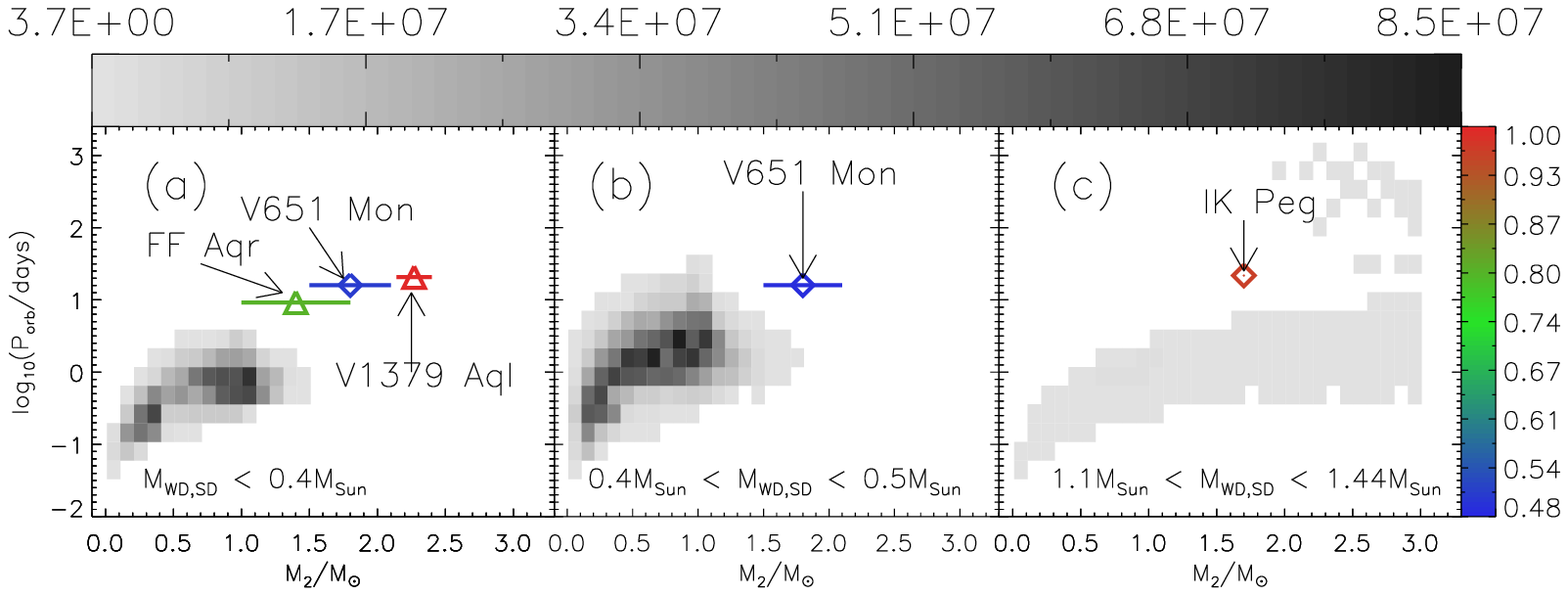}
    \caption{The theoretical present day PCEB population for
    model A, $n(q_{\rmn{i}})=1$, for a wider range in $M_{2}$ and
    $P_{\rmn{orb}}$ than Fig. \ref{multiplot}, for selected white
    dwarf mass intervals. Each panel labelled (a) to (c) represents
    the indicated range in white dwarf or sub-dwarf mass. Diamonds:
    WD+MS systems; triangles: sd+MS systems. The location of the
    individual systems V651 Mon, V1379 Aql and FF Aqr are indicated by
    the arrows. The grey-scale on the top of the plot indicates the
    number of systems per bin area, while the bar on the right hand
    side indicates the weighting of the systems in their corresponding
    panels.}
    \label{outliers}
  \end{minipage}
\end{figure*}

Both the theoretical distributions and observed systems are divided
into panels according to their white dwarf masses as indicated in each
of the nine panels, labelled (a) to (i). The grey-scale bar at the top
of the plot indicates the number of systems per bin area. We take into
account the uncertainty of the white dwarf masses for each observed
system by calculating the weighting of the system in each panel that
overlaps with the white dwarf uncertainty interval. The panels in
which these systems lie, and the associated weightings, are also shown
in Table \ref{table01}. The weightings are calculated from a Gaussian
distribution with a mean white dwarf mass
$\langle{M_{\rmn{WD}}}\rangle$ and a standard deviation $\sigma$,
which corresponds to the measured white dwarf mass and its uncertainty
respectively. The colour bar on the right hand side of the plot
indicates the weightings.

There is an acceptable overall agreement between the observed systems
and the theoretical distributions, in the sense that all but two
observed systems are in areas populated by the standard
model. However, the model distributions extend to areas at orbital
periods longer than about 1 d and donor masses larger than about 0.5
M$_{\odot}$ with very few, if any, observed systems. The two outliers
that cannot be accounted for by our standard model are V651 Mon
(Mendez \& Niemela 1981; Mendez et al. 1985; Kato, Nogami \& Baba
2001) and IK Peg (Vennes, Christian \& Thorstensen 1998). The location
of these systems are indicated in Figure \ref{outliers} by the arrows
(these systems lie outside of the range displayed in Figure
\ref{multiplot} and so are not shown there). We shall discuss these
systems in turn to deduce whether they are in fact PCEBs.

\emph{IK Peg} contains a $1.2$ M$_{\odot}$ white dwarf, and a $1.7$
M$_{\odot}$ secondary star with spectral type A8 (Landsman, Simon \&
Bergeron 1993; Smalley et al. 1996; Vennes, Christian \& Thorstensen
1998). \citet{smalley96} found that the secondary star had an
overabundance of iron, barium and strontium, which may be accounted
for if IK Peg underwent mass transfer. \citet{smalley96} suggest that
this was in the form of a CE phase. As the secondary plunged into the
envelope of the giant it would be contaminated by s-processed
material. As WD+MS systems can also form through a thermally unstable
Roche lobe overflow (RLOF) phase (see Willems \& Kolb 2004, formation
channels 1 to 3), we calculated the present day population of WD+MS
systems with $1.1\le{M_{\rmn{WD}}/\rmn{M}_{\odot}}\le{1.44}$ that form
through such a case B thermal--timescale mass transfer (TTMT) phase
(channel 2 in Willems \& Kolb 2004), which is shown in the right panel
of Figure \ref{BRLOF}. The diamond indicates the location of IK
Peg. As the location of IK Peg cannot be accounted for by our
theoretical population of case B TTMT systems, it is hence a likely
PCEB candidate. We explore this possibility further in Section 3.5
where we consider more population synthesis models of the CE phase.

\emph{V651 Mon} contains a 0.4 M$_{\odot}$ hot white dwarf primary
with an A-type secondary star. This system lies within a planetary
nebula, which could be the remnant of a CE phase. However,
\citet{dr93} suggest that the planetary nebula may have formed when
the compact primary underwent a shell flash, and instead the binary
formed via a thermally unstable RLOF phase. As for IK Peg, we
calculate the present day population of WD+MS systems with
$M_{\rmn{WD}}/\rmn{M}_{\odot}\le{0.4}$ that formed through a thermally
unstable RLOF phase, shown in the left panel of Figure
\ref{BRLOF}. This is compared with the location of V651 Mon shown as
indicated by the arrow. From Figure \ref{BRLOF} it appears hence
feasible that V651 Mon formed from a case B RLOF phase.

To obtain the possible progenitor of V651 Mon, we first find the
calculated WD+MS configuration which formed through the case B TTMT
channel which is situated closest to the location of V651 Mon in
$(M_{\rmn{WD}},M_{2},P_{\rmn{orb}})$ space. If $\delta{M_{\rmn{WD}}}$,
$\delta{M_{2}}$ and $\delta{P_{\rmn{orb}}}$ is the difference between
the observed white dwarf mass $M_{\rmn{WD,obs}}$, secondary mass
$M_{2,\rmn{obs}}$ and orbital period $P_{\rmn{orb,obs}}$ of V651 Mon
and a calculated WD+MS binary configuration, then the normalised
distance, $\Delta$, in $(M_{\rmn{WD}},M_{2},P_{\rmn{orb}})$ space is
given by
\begin{equation}
\Delta=\left(\left\vert\frac{\delta{M_{\rmn{WD}}}}{M_{\rmn{WD,obs}}}\right\vert^{2}+\left\vert\frac{\delta{M_{2}}}{M_{2,\rmn{obs}}}\right\vert^{2}+\left\vert\frac{\delta{P_{\rmn{orb}}}}{P_{\rmn{orb,obs}}}\right\vert^{2}\right)^{1/2}
\label{Delta}
\end{equation}
Thus we require the WD+MS binary configuration that gives the smallest
value of $\Delta$. We find that $\Delta$ is minimised for
$M_{\rmn{WD}}=0.33$ M$_{\odot}$, $M_{2}=1.76$ M$_{\odot}$ and
$P_{\rmn{orb}}=15.92$ d. This system has a ZAMS progenitor with
$M_{1,\rmn{i}}=2.47$ M$_{\odot}$, $M_{2,\rmn{i}}=0.98$ M$_{\odot}$ and
$P_{\rmn{orb,i}}=2.60$ d.

While the secondary mass of this WD+MS configuration is within the
measured uncertainty of the secondary mass in V651 Mon, its white
dwarf mass is not, and in fact underestimates the white dwarf mass of
V651 Mon. \citet{te88} and \citet{te88b} suggest that V651 Mon may
have formed with enhanced mass loss from the primary star before it
filled its Roche lobe. Thus, a more massive progenitor primary than
the one found in our best-fit WD+MS configuration may have formed the
white dwarf. If enough mass is lost from the primary due to enhanced
wind losses before it fills its Roche lobe to sufficiently lower the
mass ratio, $M_{1}/M_{2}$, then the resulting mass transfer will be
dynamically stable.



\subsection{Sub-dwarf + main sequence star binaries}

One of the main formation channels of sd+MS binaries is via the CE
phase \citep{han02}; if the primary star fills its Roche lobe near the
tip of the first giant branch, and the CE is ejected from the system,
then a short-period binary with an sd star is formed once the helium
core is ignited. \citet{han02} found that the mass of the sdB star
ranges from approximately 0.32 to 0.48 $M_{\odot}$. The typical mass
for a sdB star is approximately 0.46 M$_{\odot}$. This appears to be
broadly consistent with the observed masses of sd stars (see Table
\ref{table02}). The mass of the sub-dwarf in V1379 Aql is somewhat
smaller than the calculated lower limit by \citet{han02}, while the
sub-dwarfs in UU Sge and KV Kel are more massive than the calculated
upper limit in sub-dwarf mass.

The formation of sub-dwarf stars is affected by the wind loss of the
giant's envelope, the metallicity of the primary, and the degree of
convective overshooting of the primary. A detailed consideration of
sdB binaries is beyond the scope of this study. Instead, we simply
plot the observed sample of sd+MS binaries over our calculated
distributions of the WD+MS PCEBs (see Figure~\ref{multiplot2}, for
model A and $n(q_{\rmn{i}})=1$, in the same style as
Figure~\ref{multiplot}), thereby assuming that there is little orbital
evolution of these systems by the time the sdB stars become white
dwarfs. This is justified as follows:

The evolution of sd+MS binaries will be driven by wind losses from the
subdwarf primary with a mass loss rate of approximately ${10}^{-11}$
M$_{\odot}$ yr$^{-1}$ (\citet{vink04}; \citet{unglaub08}), as well as
systemic angular momentum losses. For the observed sample of sd+MS
systems in Table \ref{table02}, $M_{2}<0.35$ M$_{\odot}$ in all cases,
and so gravitational wave radiation will therefore be the only sink of
orbital angular momentum. During the lifetime of an sdB (i.e. the core
helium burning time) of approximately few$\times{10^{8}}$ yr
\citep{hpt00}, we estimate a relative change in the binary's orbital
period of less than about 7 per cent.

\begin{figure*}
  \centering
  \begin{minipage}{175mm}
  \includegraphics[scale=0.50]{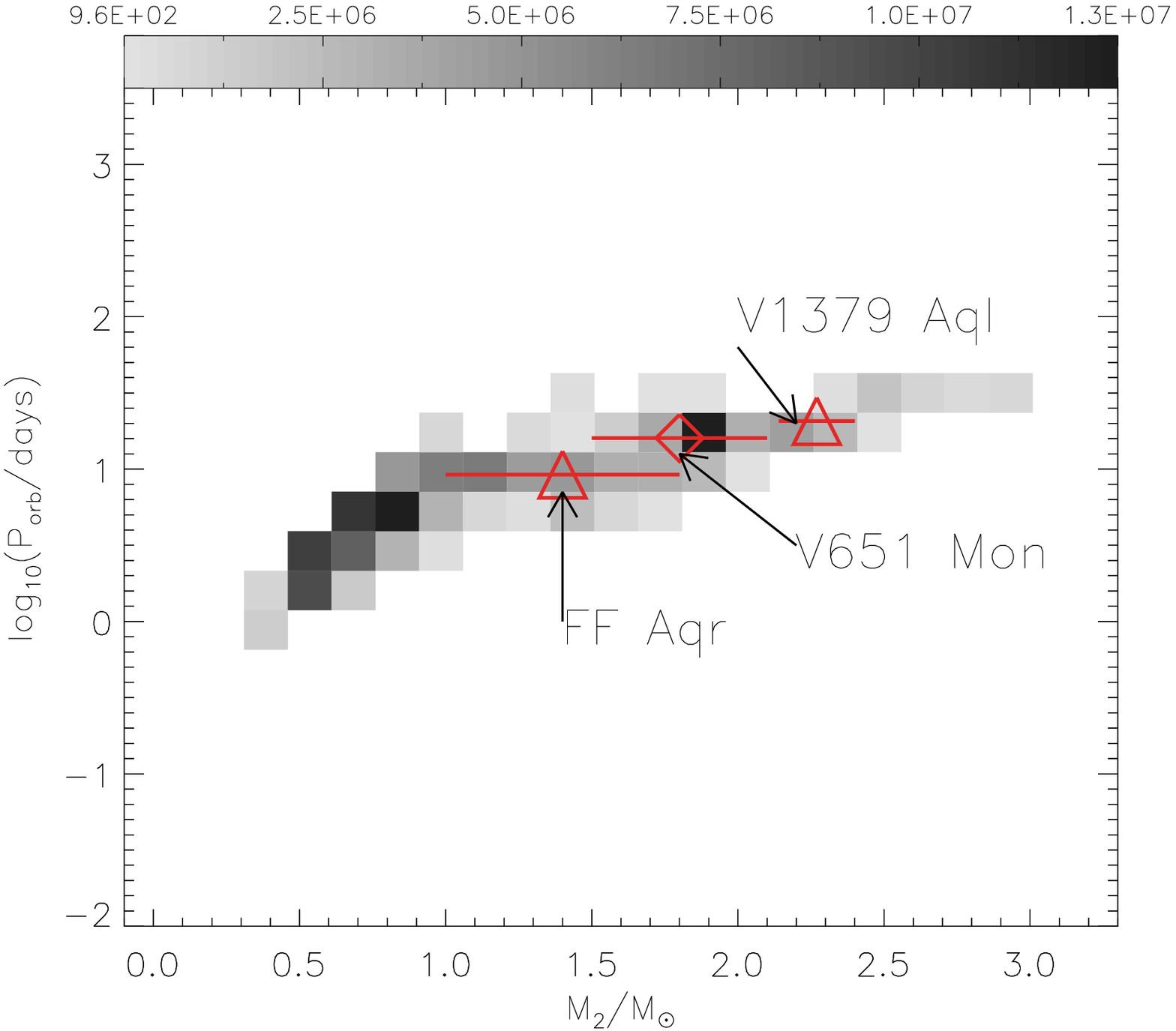}
  \includegraphics[scale=0.50]{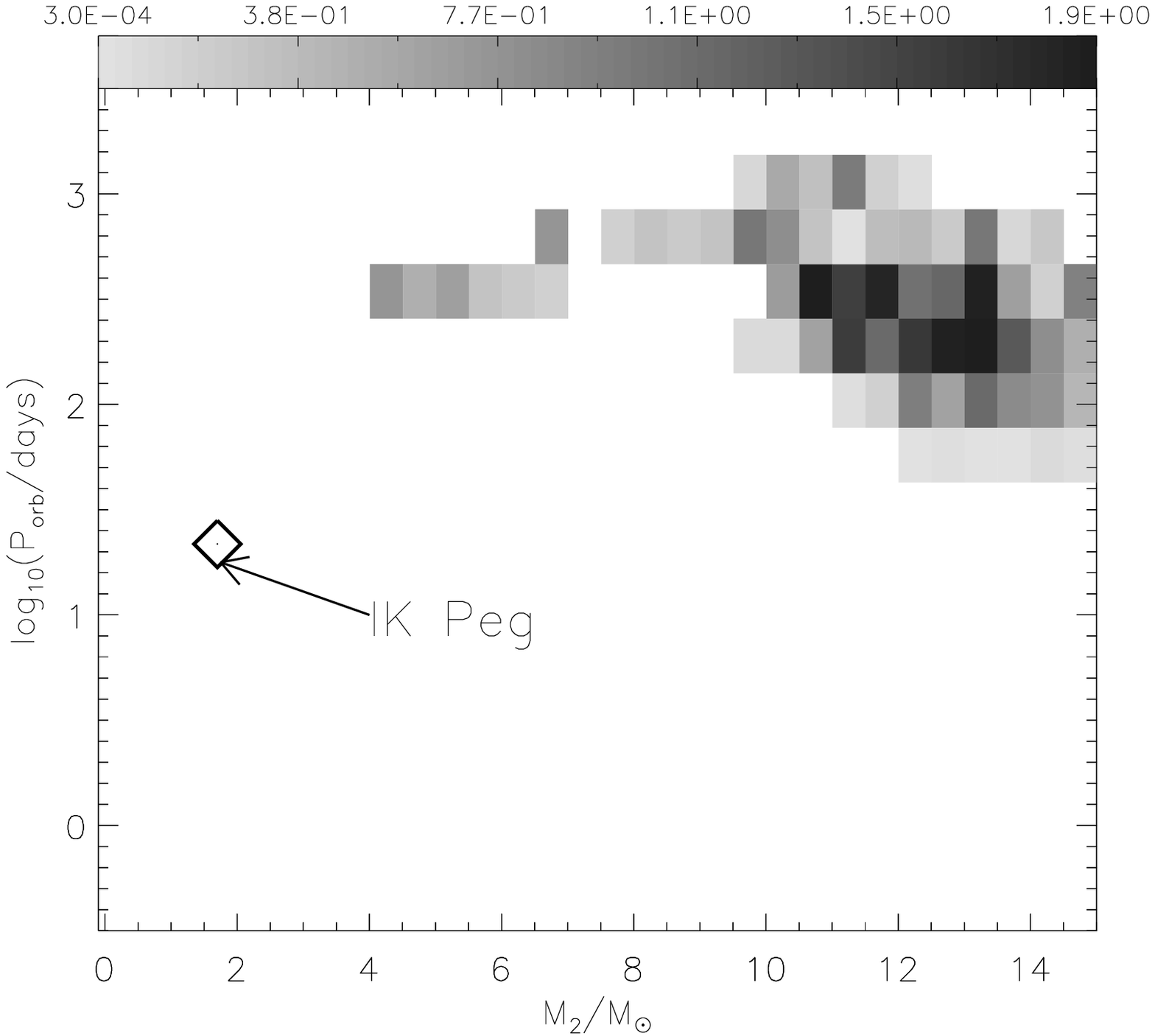}
  \caption{Left panel: The theoretical population of WD+MS
    systems with $M_{\rmn{WD}}/\rmn{M}_{\odot}\le{0.4}$ that formed
    through a thermally unstable RLOF phase, compared with the
    location of FF Aqr, V651 Mon, and V1379 Aql, as indicated by the
    arrows. Right panel: The theoretical population of WD+MS systems
    with $1.1\le{M_{\rmn{WD}}/\rmn{M}_{\odot}}\le{1.44}$ that formed
    through a thermally unstable RLOF phase with a naked helium star
    remnant. This is compared with the location of IK Peg indicated by
    the arrow. A triangular symbol indicates that the system is a
    sd+MS binary, while a diamond indicates the system is a WD+MS
    binary.}
  \label{BRLOF}
  \end{minipage}
\end{figure*}

\begin{figure*}
  \centering
  \begin{minipage}{175mm}
    \includegraphics*[viewport=18 121 492 428]{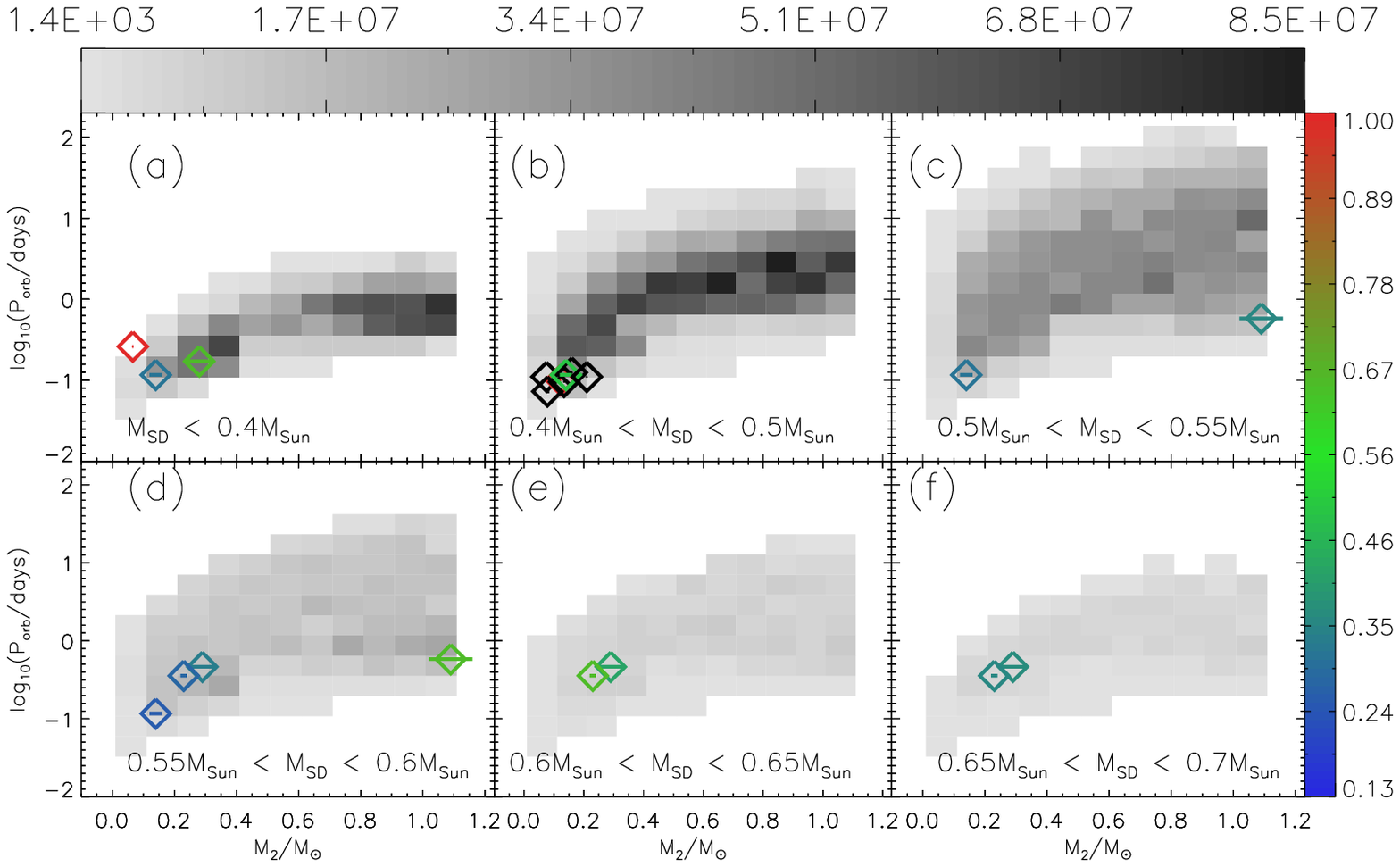}
    \caption{Same as Figure \ref{multiplot} but this time showing the
    observed sample of sub-dwarf+MS binaries. The black points
    represent those systems for which we have not calculated the
    weightings, as measured uncertainties for the sub-dwarf masses are
    not available.}
    \label{multiplot2}
  \end{minipage}
\end{figure*}

Figure \ref{multiplot2} shows that, as with the observed WD+MS
systems, there are observed sd+MS systems that cannot be explained by
our standard model. These are FF Aqr and V1379 Aql. Their locations
are indicated in the left panel of Figure \ref{outliers} by the
arrows. We discuss these systems individually to determine whether
they are in fact PCEBs.

\emph{V1379 Aql} This system contains a K-giant secondary and an sdB
primary, which may be a nascent helium white dwarf
\citep{jse92}. Furthermore, \citet{js97} argue that V1379 Aql formed
from a thermally unstable RLOF phase. As a consequence of the
synchronous rotation of the secondary, the system exhibits RS CVn
chromospheric activity due to the K-giant's enhanced magnetic
activity.  This is likely to occur if the system had a previous
episode of RLOF, rather than if the system has emerged from a CE
phase.

\citet{te88b} suggest that the progenitor binary of V1379 Aql had
$M_{1,\rmn{i}}=2.0$ M$_{\odot}$ and $M_{2,\rmn{i}}=1.6$
M$_{\odot}$. The secondary star would have completed one half of its
main sequence lifetime by the time the primary evolved to become a
giant after approximately ${6}\times{10}^{8}$ yr \citep{jse92}. If a
CE phase did occur, then the secondary would still have been on the
main sequence, and hence possess an insubstantial convective
envelope. Consequently, magnetic braking would be ineffective in
shrinking the orbital separation sufficiently for tidal interactions
to enforce synchronous rotation of the secondary star.

\citet{nt05}, in reconstructing a possible CE phase for this system,
could not find a solution for $\alpha_{\rmn{CE}}$, and cited this as
evidence for their alternative CE description in terms of the angular
momentum balance. We suggest that no solution was found because of the
possibility that V1379 Aql formed from a thermally unstable RLOF
phase.

The left panel of Figure \ref{BRLOF} compares the theoretical WD+MS
systems with $M_{\rmn{WD}}/\rmn{M}_{\odot}\le{0.4}$ that formed
through a thermally unstable RLOF phase with the location of V1379 Aql
shown by the arrow. It appears hence feasible that V1379 Aql formed
through this evolutionary channel. As with V651 Mon, we apply equation
(\ref{Delta}) to find the WD+MS configuration which lies the closest
to the location of V1379 Aql in
$(M_{\rmn{WD}},M_{2},P_{\rmn{orb}})$. Doing this we find a WD+MS
configuration with $M_{\rmn{WD}}=0.23$ M$_{\odot}$, $M_{2}=2.31$
M$_{\odot}$ and $P_{\rmn{orb}}=20.58$ d. This system forms from a ZAMS
progenitor with $M_{1,\rmn{i}}=1.79$ M$_{\odot}$, $M_{2,\rmn{i}}=0.77$
M$_{\odot}$ and $P_{\rmn{orb,i}}=1.41$ d.

The mass of the white dwarf in our best-fit WD+MS configuration once
again underestimates that of V1379 Aql. As for V651 Mon,\citet{te88b}
suggest that the primary star in V1379 Aql also underwent enhanced
mass loss before it filled its Roche lobe. Thus, the mass ratio,
$M_{1}/M_{2}$, was decreased sufficiently so that a dynamically
unstable RLOF phase was avoided. The actual progenitor primary of
V1379 Aql may therefore have been more massive than what we have
predicted in our calculations. Note also that in our calculations, we
do not treat the sdB phase of the white dwarf explicitly.

\emph{FF Aqr} also displays RS CVn characteristics \citep{vw03}, and
hence we can apply the same argument used for V1379 Aql to FF Aqr; the
secondary has a mass of approximately 1.4 M$_{\odot}$ and so will have
an insubstantial convective envelope. Indeed, \citet{dr93} argue that
FF Aqr formed from a thermally unstable RLOF phase, perhaps with
enhanced wind losses \citep{te88b}.

The left panel of Figure \ref{BRLOF}, compares the theoretical WD+MS
binary population with $M_{\rmn{WD}}/\rmn{M}_{\odot}\le{0.4}$, which
formed through a channel 1 \citet{wk04} RLOF phase, with the location
of FF Aqr indicated by the arrow. This supports the possibility that
FF Aqr formed from a TTMT RLOF phase. While our best fit WD+MS
configuration has $M_{\rmn{WD}}=0.30$ M$_{\odot}$, which is consistent
with the observed value within the measured uncertainty, our best-fit
configuration slightly underestimates the secondary mass with
$M_{2}=1.22$ M$_{\odot}$.

\subsection{The shape of the distributions}

\begin{figure}
  \includegraphics[scale=0.4]{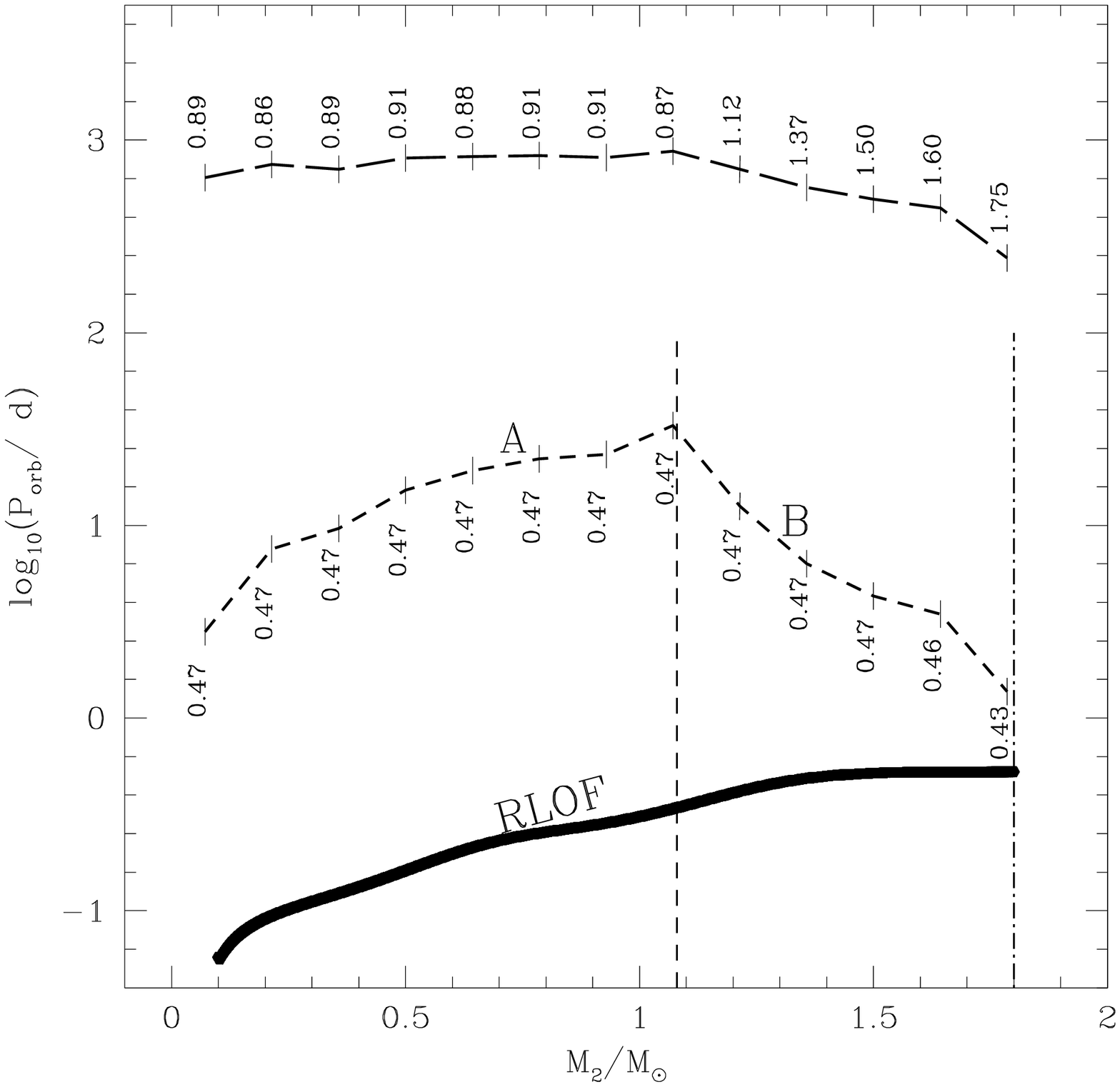}
  \caption{Critical boundaries in the
  $\rmn{log}_{10}(P_{\rmn{orb}})-M_{2}$ plane. The long dashed line
  shows the upper boundary in the population of progenitor binaries at
  the start of the CE phase which will form PCEBs with
  $0.4<M_{\rmn{WD}}/\rmn{M}_{\odot}\le{0.5}$. The values along this
  line denote the mass of the Roche lobe-filling progenitor primary 
  at the corresponding locations along the boundary. The short dashed
  line is the upper boundary in the resulting population of PCEBs with
  $0.4<M_{\rmn{WD}}\le{0.5}$. The values along this line denote the
  white dwarf mass 
  The label `A' corresponds to the portion of the boundary
  at $M_{2}\la{1.1}$ M$_{\odot}$ (short-dashed line), while `B'
  denotes the portion of the boundary at $M_{2}\ga{1.1}$
  M$_{\odot}$. The thick black line is the Roche lobe-overflow (RLOF)
  limit, i.e. the point at which the PCEB will become semi-detached,
  and hence defines the lower limit in the distribution. Finally, the
  dot-dashed line corresponds to the cut-off in this PCEB population
  at $M_{2}\approx{1.8}$ M$_{\odot}$.}
  \label{boundaries}
\end{figure}

We will now discuss common features of the
$M_{2}-\rmn{log_{10}}P_{\rmn{orb}}$ PCEB distributions, with the
middle panel of Figure \ref{outliers} as a typical
example. \footnote{We note that the shapes of the distributions for
model n15 are very different. We refer the reader to \citet{nt05} for
a detailed discussion.}

Figure \ref{boundaries} shows the upper boundary in the population of
progenitor systems at the onset of the CE phase which will form PCEBs
with $0.4<M_{\rmn{WD}}/\rmn{M}_{\odot}\le{0.5}$ (long dashed
line). The short dashed line shows the corresponding upper boundary in
the resulting population of PCEBs (this represents the upper boundary
of the distribution shown in the middle panel of Figure
\ref{outliers}), while the thick, black line shows the orbital period
at which the secondary star will undergo Roche lobe overflow (RLOF),
and hence the PCEB will become semi-detached.

The long dashed boundary in Figure \ref{boundaries} arises because for
each mass $M_2$ there is an upper limit for the orbital period of the
progenitor binary that will still form a white dwarf with a mass
within the considered mass interval. Longer-period systems will either
form a more massive white dwarf, or remain detached and not undergo a
CE phase. More specifically, the upper boundary corresponds to the
most massive white dwarfs in the panel. In the example shown this is
0.47 M$_{\odot}$ (not 0.50 M$_{\odot}$) because there is a gap in the
white dwarf mass spectrum between the low-mass helium white dwarfs and
the more massive carbon-oxygen white dwarfs. The SSE code
\citep{hpt00} places this gap in the range between
$0.47<M_{\rmn{WD}}/\rmn{M}_{\odot}<0.6$.

This long dashed upper pre-CE boundary maps on to the short dashed
PCEB boundary, also shown in Figure \ref{boundaries}. The masses of
the progenitor primaries at the start of the CE phase are indicated
along the long-dashed boundary, while the white dwarf masses are shown
along the short-dashed boundary of the PCEB population, at the
corresponding point along the boundary.

For a PCEB to form, the ZAMS progenitor primary must fill its Roche
lobe and commence the CE phase before the secondary itself evolves off
the main sequence. This point is of little consequence for
$M_{2,\rmn{i}}\la{1}$ M$_{\odot}$ as the main sequence lifetime of
such secondaries will be $\tau_{\rmn{MS,2}}\ga{10}$ Gyr, which is on
the order of the Galactic lifetime (and the maximum evolution time we
consider). Thus, in such cases, the progenitor primary must ascend the
giant branch and fill its Roche lobe within the lifetime of the
Galaxy.

We find that the least massive ZAMS progenitor which will subsequently
form a 0.47 M$_{\odot}$ white dwarf within the lifetime of the Galaxy,
is 1.1 M$_{\odot}$. Hence, for the reasons stated above,
$M_{2,\rmn{i}}<1.1$ M$_{\odot}$, if the evolution of the secondary off
the main sequence is to remain unimportant. As indicated along the
long-dashed boundary in Figure \ref{boundaries} for $M_{2}<1.1$
M$_{\odot}$(to the left of the dashed vertical line), the progenitor
primaries, as a result of wind losses, commence the CE phase with
masses of approximately 0.9 M$_{\odot}$. Furthermore, as the
long-dashed boundary for $M_{2}<1.1$ M$_{\odot}$ is approximately
flat, the progenitor primaries along this boundary fill their Roche
lobes at the same orbital period.

To see how this determines the shape of the resulting short-dashed
PCEB boundary for $M_{2}<1.1$ M$_{\odot}$ (labelled `A' in Figure
\ref{boundaries}) we consider equation (\ref{af}), which shows that
for constant $M_{\rmn{c}}$ (and hence the white dwarf mass), $M_{1}$
and $r_{\rmn{L,1}}$ (at the start of the CE phase), $A_{\rmn{CE,f}}$
decreases for decreasing $M_{2}$. Hence the orbital period of the
short-dashed boundary to the left of the dashed line in Figure
\ref{boundaries} decreases for decreasing $M_{2}$.

On the other hand, for $M_{2,\rmn{i}}>1.1$ M$_{\odot}$ (to the right
of the dashed line in Figure \ref{boundaries}), the main sequence
lifetime of such stars is less than 10 Gyr, and so the progenitor
primary must increasingly compete against the secondary to evolve off
the main sequence first. As a result, the ZAMS progenitor primary star
needs to be increasingly more massive for increasing secondary
mass. This is reflected in the masses of the primaries at the start of
the CE phase, as indicated along the long-dashed boundary in Figure
\ref{boundaries} for $M_{2}>1.1$ M$_{\odot}$, which increase from
approximately 0.9 M$_{\odot}$ where $M_{2}\approx{1.1}$ M$_{\odot}$,
to 1.8 M$_{\odot}$ where $M_{2}\approx{1.8}$ M$_{\odot}$.

For a primary star ascending the giant branch with luminosity $L_{1}$,
its radius on the giant branch, $R_{\rmn{GB}}$, can be modelled as
\begin{equation}
R_{\rmn{GB}}\approx\frac{1.1}{M_{1,\rmn{i}}^{0.3}}\left(L_{1}^{0.4}+0.383L_{1}^{0.76}\right),
\label{Rgb}
\end{equation}
\citep{hpt00}, where $L_{1}\propto{M_{\rmn{c}}}^{6}$. Thus the radius
of the star on the giant branch is a function of its core mass and
(weakly) of its initial ZAMS mass. For a given core mass, in this case
of $M_{\rmn{c}}=0.47$ M$_{\odot}$, the corresponding value of
$R_{\rmn{GB}}$ will decrease for increasing mass of the ZAMS
primary. For Roche lobe filling stars we have
$P_{\rmn{orb}}\propto{R}^{3/2}/M^{1/2}$ so the orbital period at which
the CE phase will occur will decrease with increasing $M_{2}$. This is
shown as the slope in the long-dashed boundary in Figure
\ref{boundaries} for $M_{2}>1.1$ M$_{\odot}$.

The shape of the short-dashed boundary to the right of the dashed line
in Figure \ref{boundaries} (labelled `B') is a consequence of the
shape of the corresponding portion of the long-dashed pre-CE
boundary. Furthermore, for increasing $M_{2}$, while M$_{\rmn{c}}$
remains constant, $M_{\rmn{1,i}}$ does increase, and so the mass of
the primary giant's envelope will correspondingly increase. The result
of this is that $A_{\rmn{CE,f}}$ will decrease for increasing $M_{2}$,
as equation (\ref{af}) shows. This explains why the short-dashed PCEB
boundary to the right of the dashed lin Figure \ref{boundaries}
decreasing for increasing $M_{2}$.

Note that the cut-off in the population, as indicated by the vertical
dot-dashed line (at $M_{2}\approx{1.8}$ M$_{\odot}$) in Figure
\ref{boundaries}, is a consequence of the fact that the most massive
progenitor primary which will form a white dwarf in the range
$0.4<M_{\rmn{WD}}/\rmn{M}_{\odot}\le{0.5}$ is 1.8 M$_{\odot}$. For
PCEB distributions with larger white dwarf masses, which will be
formed from more massive primary progenitors, this cut-off will shift
towards larger values of $M_{2}$.

\begin{figure*}
  \begin{minipage}{175mm}
    \includegraphics[scale=0.50]{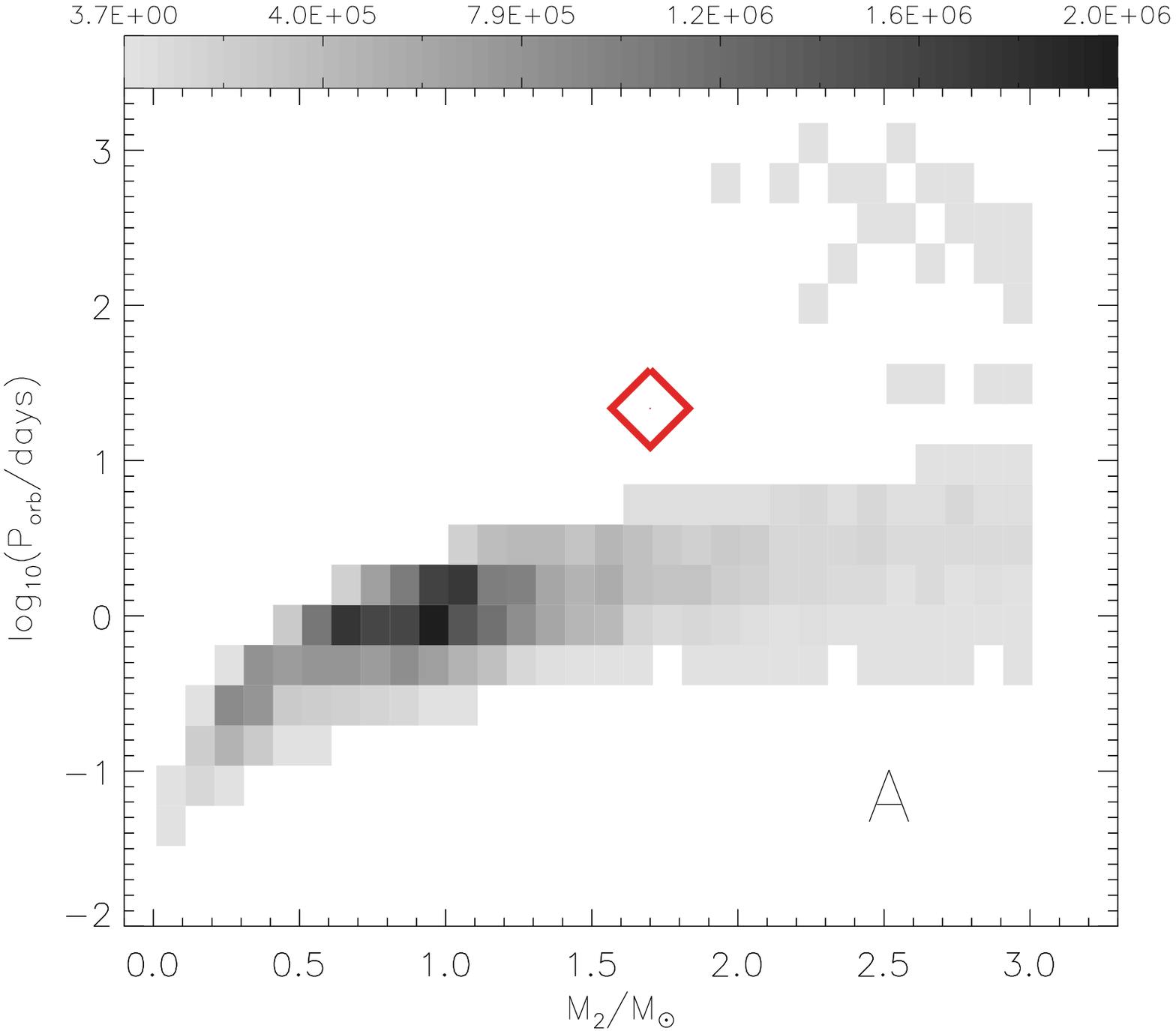}
    \includegraphics[scale=0.50]{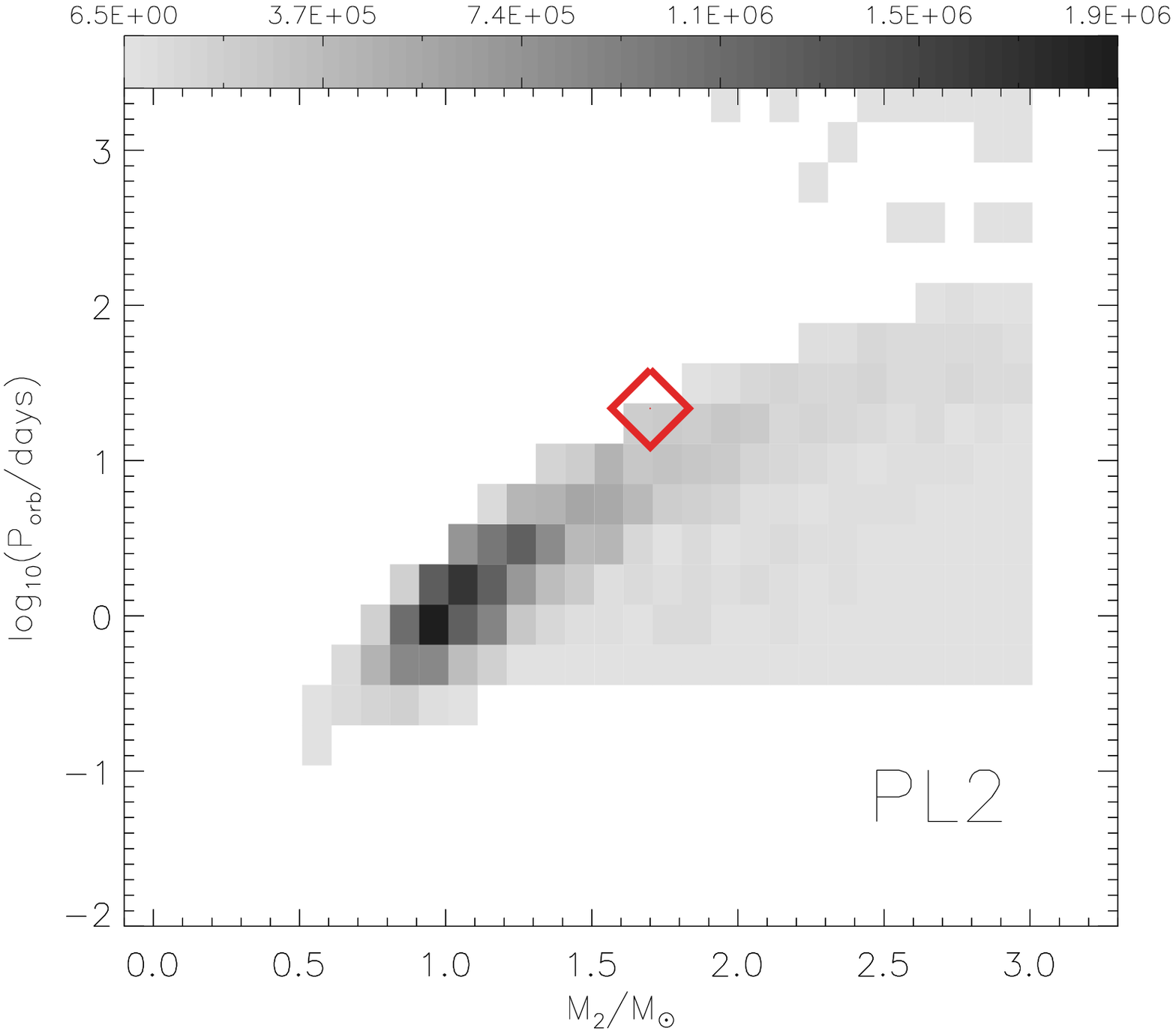}
    \includegraphics[scale=0.50]{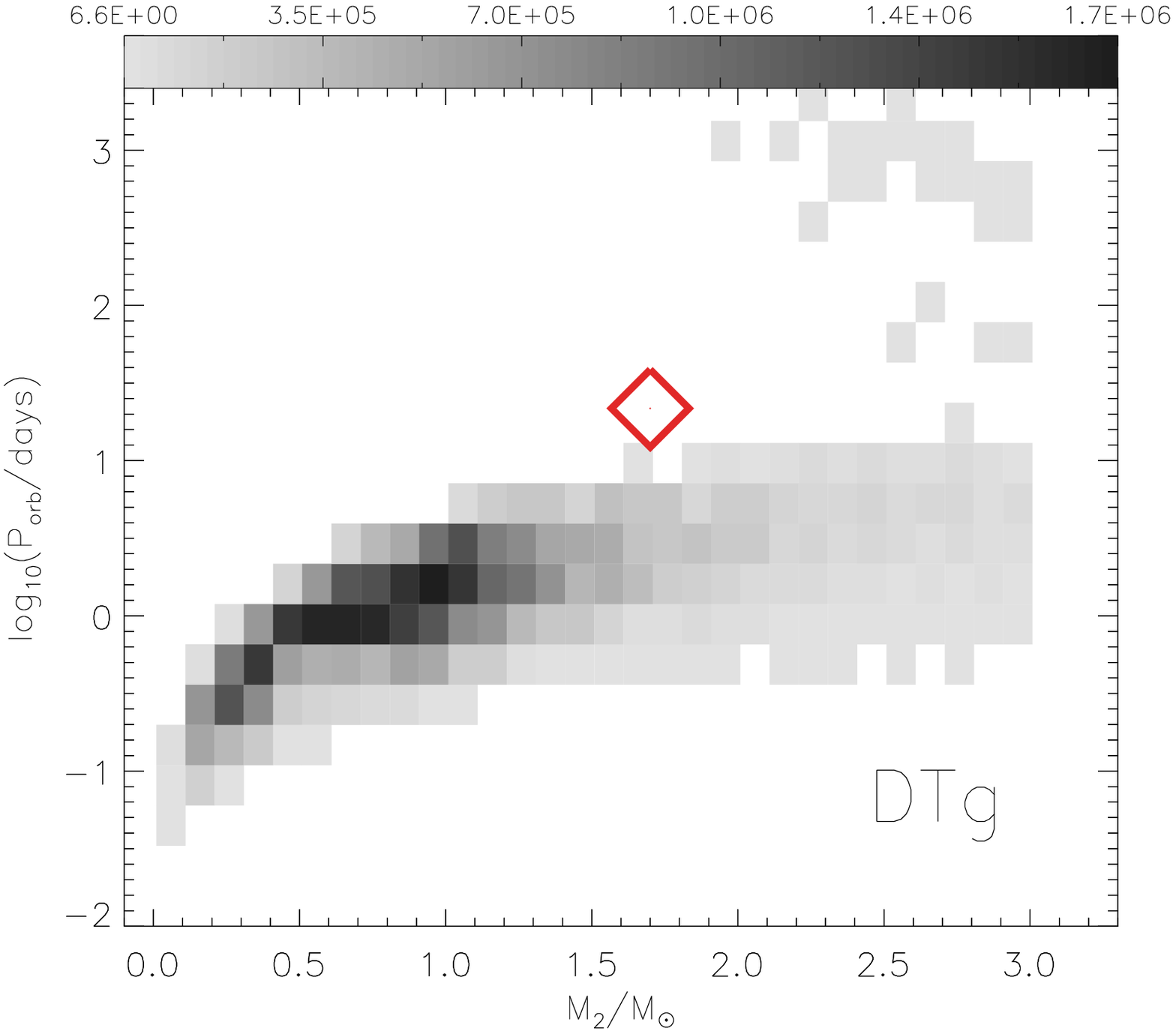}
    \includegraphics[scale=0.50]{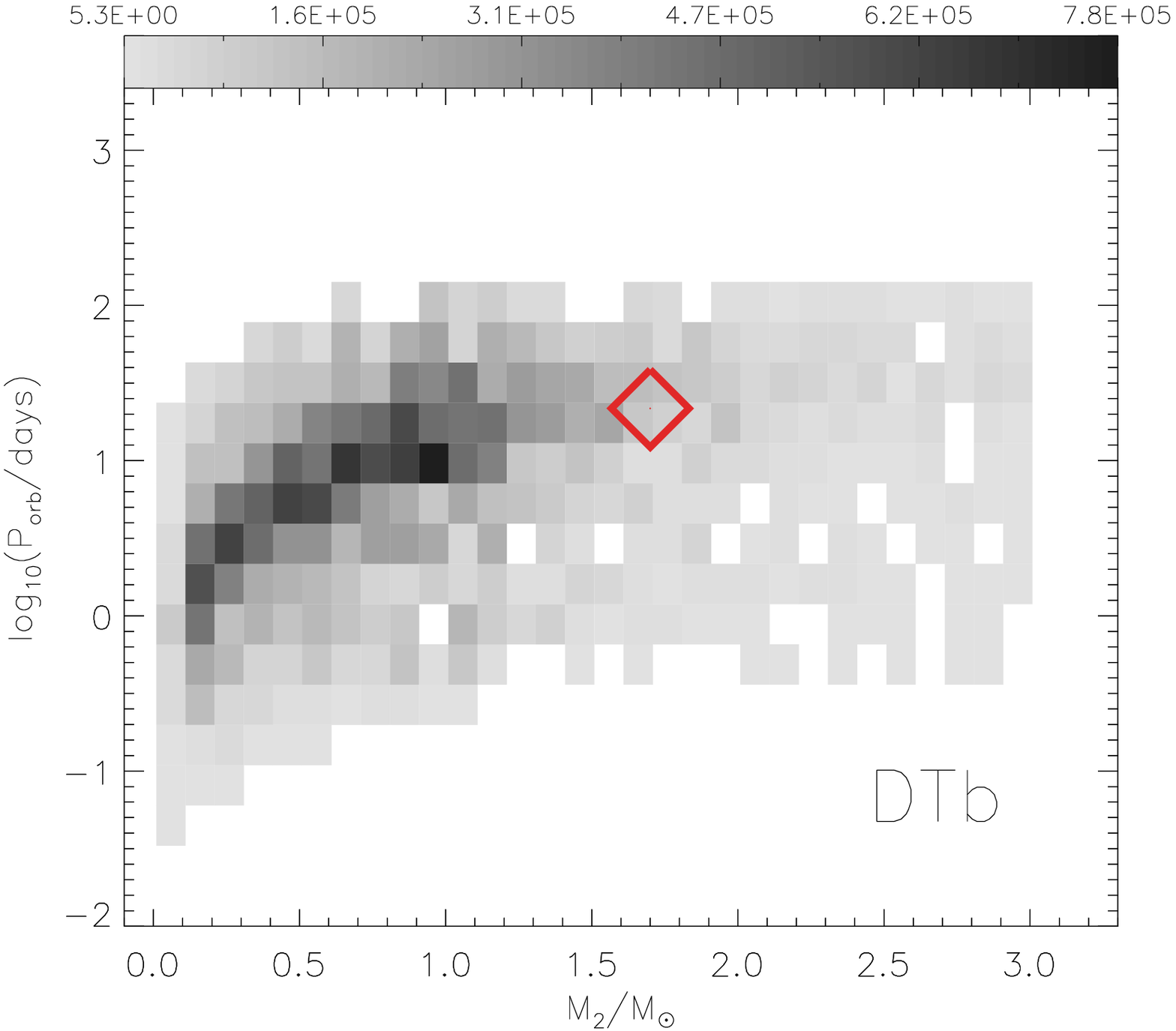}
    \begin{center}
    \includegraphics[scale=0.50]{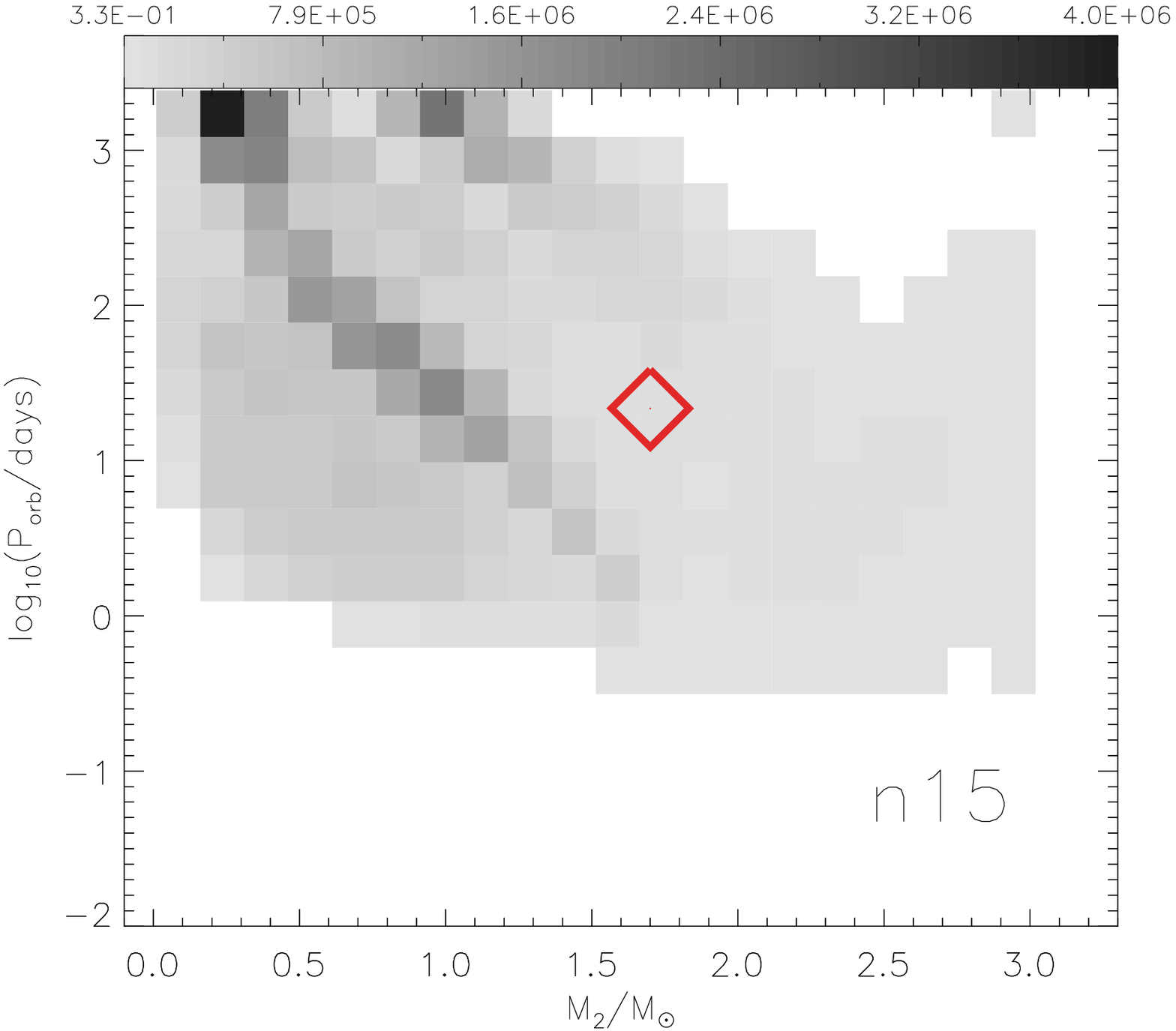}
    \end{center}
    \caption{Calculated PCEB populations with
    $1.1<M_{\rmn{WD}}/\rmn{M}_{\odot}\le{1.44}$ for a range of models
    indicated in the lower right hand corner of each plot. Here
    $n(q_{\rmn{i}})=1$. The observed location of IK Peg is indicated
    by a diamond. The grey-scale bars above each plot indicates the
    number of systems per bin area.}
    \label{multiplot3}
  \end{minipage}
\end{figure*}

\subsection{The present day population and space densities of PCEBs}

Table \ref{table_stats} summarises the formation rate, present day
population and the local space density, $\varrho$, of PCEBs for each
model describing the treatment of the CE phase, and for each
population model we considered. We calculate the local space density
of PCEBs by dividing the present day population by the Galactic volume
of $5\times{10}^{11}$ pc$^{3}$.

For a given initial secondary mass distribution, we see that the
present day population of PCEBs increases with increasing values of
$\alpha_{\rmn{CE}}$. By increasing $\alpha_{\rmn{CE}}$, less orbital
energy is required to eject the CE and therefore the binary system
will undergo less spiral in. Hence, fewer systems will merge during
the CE phase. For a given IMRD, we find very little difference in the
calculated formation rates, present day numbers and local space
densities of PCEBs between models A and DTg. For model DTb, on the
other hand, where we are now considering the internal energy of the
giant's envelope, we obtain a modest increase in the present day
population of PCEBs compared to model A.

The present day population of PCEBs decreases slightly with increasing
values of $p$ in equation (\ref{alpha_func_m2}). As shown in Figure 1
of \citet{pw07} the average value of $\alpha_{\rmn{CE}}$ for a given
range in $M_{2}$ decreases with increasing $p$. By increasing $p$ more
orbital energy is required to expel the envelope from the system,
hence a larger spiral-in of the binary. This will in turn lead to more
mergers.

On the other hand, the present day number (local space density) of
PCEBs decreases from model CT0375 to model CT15. As we increase the
value of $M_{\rmn{cut}}$ the typical value of $\alpha_{\rmn{CE}}$ for
a given range of $M_{2}$ will decrease. Hence, fewer systems will
survive the CE phase.

We notice that the model n15 which describes the CE phase in terms of
the binary's angular momentum gives the largest present day population
of PCEBs. For $n(q_{\rmn{i}})=1$, the present day number (local space density)
of PCEBs is $6.1\times{10}^{8}$ ($1.2\times{10}^{-3}$ pc$^{-3}$).

\citet{sg03} estimated from the observed sample of PCEBs that
$6\times{10}^{-6}\la{\varrho}/\rmn{pc}^{-3}\la{3\times{10}^{-5}}$. An
IMRD of $n(q_{\rmn{i}})\propto{q_{\rmn{i}}}^{-0.99}$ gives the best
agreement with this observed estimate, where we obtain
$6.0\times{10}^{-7}\le{\varrho}/\rmn{pc}^{-3}\le{5.6\times{10}^{-5}}$.

Our models with $\alpha_{\rmn{CE}}=0.1$ also provide small space
densities
($6.0\times{10}^{-7}\le{\varrho/\rmn{pc}^{-3}}\le{1.7\times{10}^{-5}}$,
depending on the IMRD), which are similar to the observed
ones. However, we find that in these models \emph{no} PCEBs with
M$_{\rmn{WD}}/\rmn{M_{\odot}}\le{0.4}$ form. Low mass helium white
dwarfs form via a case B CE phase, when the primary fills its Roche
lobe on the first giant branch, typically when $R_{1}\approx{15}$ to
100 R$_{\odot}$. This means that the orbital period at the onset of
the CE phase is between approximately 40 and 250 d. As a consequence
of such a low ejection efficiency, the binary cannot spiral in
sufficiently to eject the envelope before a merger occurs. The lack of
PCEBs with low mass white dwarfs in models with
$\alpha_{\rmn{CE}}=0.1$ is in conflict with observations; for example
LM Com and 0137-3457 have 95 per cent and 61 per cent probabilities
respectively of having $M_{\rmn{WD}}/\rmn{M}_{\odot}\le{0.4}$.

\begin{table}
  \centering
    \caption{The formation rates, present day numbers and local space
    densities of PCEBs for different treatments of the CE phase, and
    for different initial mass distributions of the secondary
    star. The space densities were calculated by dividing the present
    day numbers by the Galactic volume of $5\times{10}^{11}$
    pc$^{3}$.}
    \begin{tabular}{@{}lccc@{}}
      \hline
      Model         &          Formation rate         &       Number of systems         &         Local space    \\
                    &            /yr$^{-1}$       &                                     &       density  /pc$^{-3}$        \\
      \hline

      \multicolumn{4}{c}{$n(q_{\rmn{i}})\propto{q_{\rmn{i}}^{-0.99}}$,
      $0<q_{\rmn{i}}\le{1}$} \\

      \hline
      CE01         &          $2.1\times{10}^{-4}$     &      $3.0\times{10}^{5}$        &        $6.0\times{10}^{-7}$     \\
      CE06         &          $1.5\times{10}^{-3}$     &      $4.4\times{10}^{6}$        &        $8.8\times{10}^{-6}$     \\
      A            &          $2.0\times{10}^{-3}$     &      $7.3\times{10}^{6}$        &        $1.5\times{10}^{-5}$     \\
      \hline
      PL05         &          $1.6\times{10}^{-3}$     &      $4.8\times{10}^{6}$        &        $9.6\times{10}^{-6}$     \\
      PL1          &          $1.3\times{10}^{-3}$     &      $3.6\times{10}^{6}$        &        $7.2\times{10}^{-6}$     \\
      PL2          &          $1.1\times{10}^{-3}$     &      $2.7\times{10}^{6}$        &        $5.4\times{10}^{-6}$     \\
      \hline
      CT0375       &          $1.8\times{10}^{-3}$     &      $6.6\times{10}^{6}$        &        $1.3\times{10}^{-5}$     \\
      CT075        &          $1.7\times{10}^{-3}$     &      $5.7\times{10}^{6}$        &        $1.1\times{10}^{-5}$     \\
      CT15         &          $1.3\times{10}^{-3}$     &      $4.2\times{10}^{6}$        &        $8.5\times{10}^{-6}$     \\
      \hline
      DTg          &          $1.9\times{10}^{-3}$     &      $6.8\times{10}^{6}$        &        $1.3\times{10}^{-5}$     \\
      DTb          &          $2.5\times{10}^{-3}$     &      $1.2\times{10}^{7}$        &        $2.4\times{10}^{-5}$     \\
      \hline
      n15          &          $5.4\times{10}^{-3}$     &      $2.8\times{10}^{7}$        &        $5.6\times{10}^{-5}$     \\
      \hline

      \multicolumn{4}{c}{$n(q_{\rmn{i}})=1$, $0<q_{\rmn{i}}\le{1}$ (our reference IMRD)}  \\

      \hline
      CE01        &          $9.8\times{10}^{-3}$      &      $1.5\times{10}^{7}$        &        $3.0\times{10}^{-5}$     \\
      CE06        &          $5.5\times{10}^{-2}$      &      $1.5\times{10}^{8}$        &        $3.0\times{10}^{-4}$     \\
      A           &          $7.2\times{10}^{-2}$      &      $2.2\times{10}^{8}$        &        $4.4\times{10}^{-4}$     \\
      \hline
      PL05        &          $6.8\times{10}^{-2}$      &      $1.9\times{10}^{8}$        &        $3.8\times{10}^{-4}$     \\
      PL1         &          $6.5\times{10}^{-2}$      &      $1.6\times{10}^{8}$        &        $3.2\times{10}^{-4}$     \\
      PL2         &          $6.1\times{10}^{-2}$      &      $1.4\times{10}^{8}$        &        $2.8\times{10}^{-4}$     \\
      \hline
      CT0375      &          $7.0\times{10}^{-2}$      &      $2.1\times{10}^{8}$        &        $4.3\times{10}^{-4}$     \\
      CT075       &          $6.7\times{10}^{-2}$      &      $2.0\times{10}^{8}$        &        $4.0\times{10}^{-4}$     \\
      CT15        &          $6.0\times{10}^{-2}$      &      $1.7\times{10}^{8}$        &        $3.5\times{10}^{-4}$     \\
      \hline
      DTg         &          $7.0\times{10}^{-2}$      &      $2.0\times{10}^{8}$        &        $4.0\times{10}^{-4}$     \\
      DTb         &          $8.7\times{10}^{-2}$      &      $3.1\times{10}^{8}$        &        $6.2\times{10}^{-4}$     \\
      \hline
      n15         &          $3.6\times{10}^{-1}$      &      $6.1\times{10}^{8}$        &        $1.2\times{10}^{-3}$     \\
      \hline

      \multicolumn{4}{c}{$n(q_{\rmn{i}})\propto{q_{\rmn{i}}}$, $0<q_{\rmn{i}}\le{1}$}\\

      \hline
      CE01        &          $1.1\times{10}^{-2}$      &      $1.9\times{10}^{7}$        &        $3.8\times{10}^{-5}$     \\
      CE06        &          $5.8\times{10}^{-2}$      &      $1.4\times{10}^{8}$        &        $2.8\times{10}^{-4}$     \\
      A           &          $7.5\times{10}^{-2}$      &      $2.0\times{10}^{8}$        &        $4.0\times{10}^{-4}$     \\
      \hline
      PL05        &          $7.7\times{10}^{-2}$      &      $1.8\times{10}^{8}$        &        $3.6\times{10}^{-4}$  \\
      PL1         &          $7.8\times{10}^{-2}$      &      $1.8\times{10}^{8}$        &        $3.6\times{10}^{-4}$  \\
      PL2         &          $7.8\times{10}^{-2}$      &      $1.6\times{10}^{8}$        &        $3.2\times{10}^{-4}$  \\
      \hline
      CT0375      &          $7.3\times{10}^{-2}$      &      $2.0\times{10}^{8}$        &        $3.9\times{10}^{-4}$  \\
      CT075       &          $7.2\times{10}^{-2}$      &      $1.9\times{10}^{8}$        &        $3.8\times{10}^{-4}$  \\
      CT15        &          $6.8\times{10}^{-2}$      &      $1.7\times{10}^{8}$        &        $3.5\times{10}^{-4}$  \\
      \hline
      DTg         &          $7.3\times{10}^{-2}$      &      $1.8\times{10}^{8}$        &        $3.6\times{10}^{-4}$  \\
      DTb         &          $9.0\times{10}^{-2}$      &      $2.7\times{10}^{8}$        &        $5.4\times{10}^{-4}$  \\
      \hline
      n15         &          $1.7\times{10}^{-1}$      &      $4.9\times{10}^{8}$        &        $9.8\times{10}^{-4}$  \\
      \hline

      \multicolumn{4}{c}{IMFM2}\\

      \hline
      CE01       &           $6.5\times{10}^{-3}$      &      $8.3\times{10}^{6}$        &        $1.7\times{10}^{-5}$  \\
      CE06       &           $4.7\times{10}^{-2}$      &      $1.5\times{10}^{8}$        &        $3.0\times{10}^{-4}$  \\
      A          &           $6.5\times{10}^{-2}$      &      $2.5\times{10}^{8}$        &        $5.0\times{10}^{-4}$  \\
      \hline
      PL05       &           $4.4\times{10}^{-2}$      &      $1.4\times{10}^{8}$        &        $2.8\times{10}^{-4}$  \\
      PL1        &           $3.0\times{10}^{-2}$      &      $8.3\times{10}^{7}$        &        $1.7\times{10}^{-4}$  \\
      PL2        &           $1.7\times{10}^{-2}$      &      $4.7\times{10}^{7}$        &        $9.4\times{10}^{-5}$  \\
      \hline
      CT0375     &           $5.8\times{10}^{-2}$      &      $2.2\times{10}^{8}$        &        $4.4\times{10}^{-4}$  \\
      CT075      &           $4.9\times{10}^{-2}$      &      $1.8\times{10}^{8}$        &        $3.5\times{10}^{-4}$  \\
      CT15       &           $3.3\times{10}^{-2}$      &      $1.1\times{10}^{8}$        &        $2.2\times{10}^{-4}$  \\
      \hline
      DTg        &           $6.1\times{10}^{-2}$      &      $2.3\times{10}^{8}$        &        $4.6\times{10}^{-4}$  \\
      DTb        &           $8.5\times{10}^{-2}$      &      $4.2\times{10}^{8}$        &        $8.4\times{10}^{-4}$  \\
      \hline
      n15        &           $1.9\times{10}^{-1}$      &      $1.1\times{10}^{9}$        &        $2.2\times{10}^{-3}$  \\
      \hline
      \label{table_stats}

    \end{tabular}
\end{table}

\subsection{IK Peg: A Clue to the CE Mechanism?}

\begin{figure}
  \centering
    \includegraphics[scale=0.35]{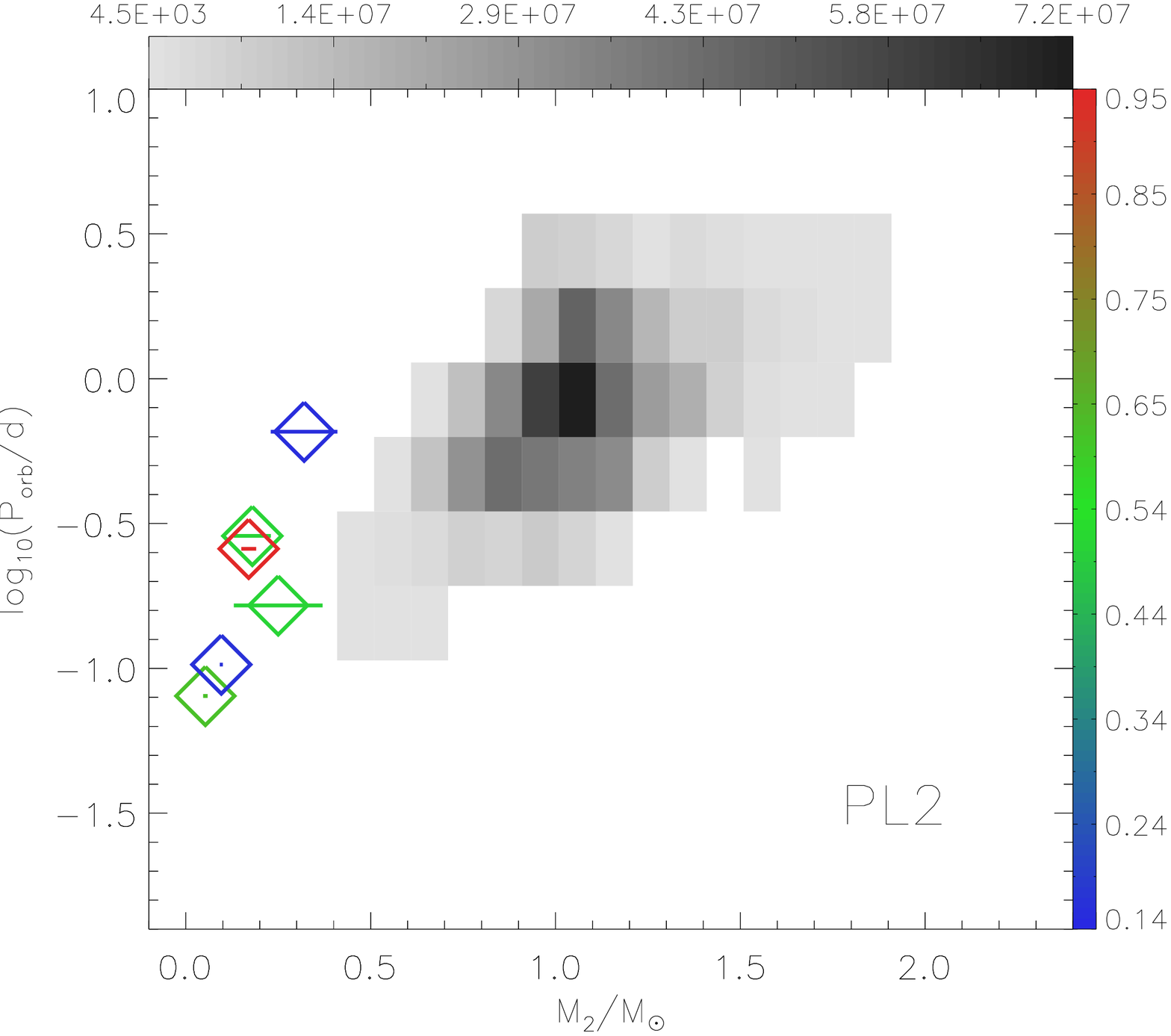}
    \includegraphics[scale=0.35]{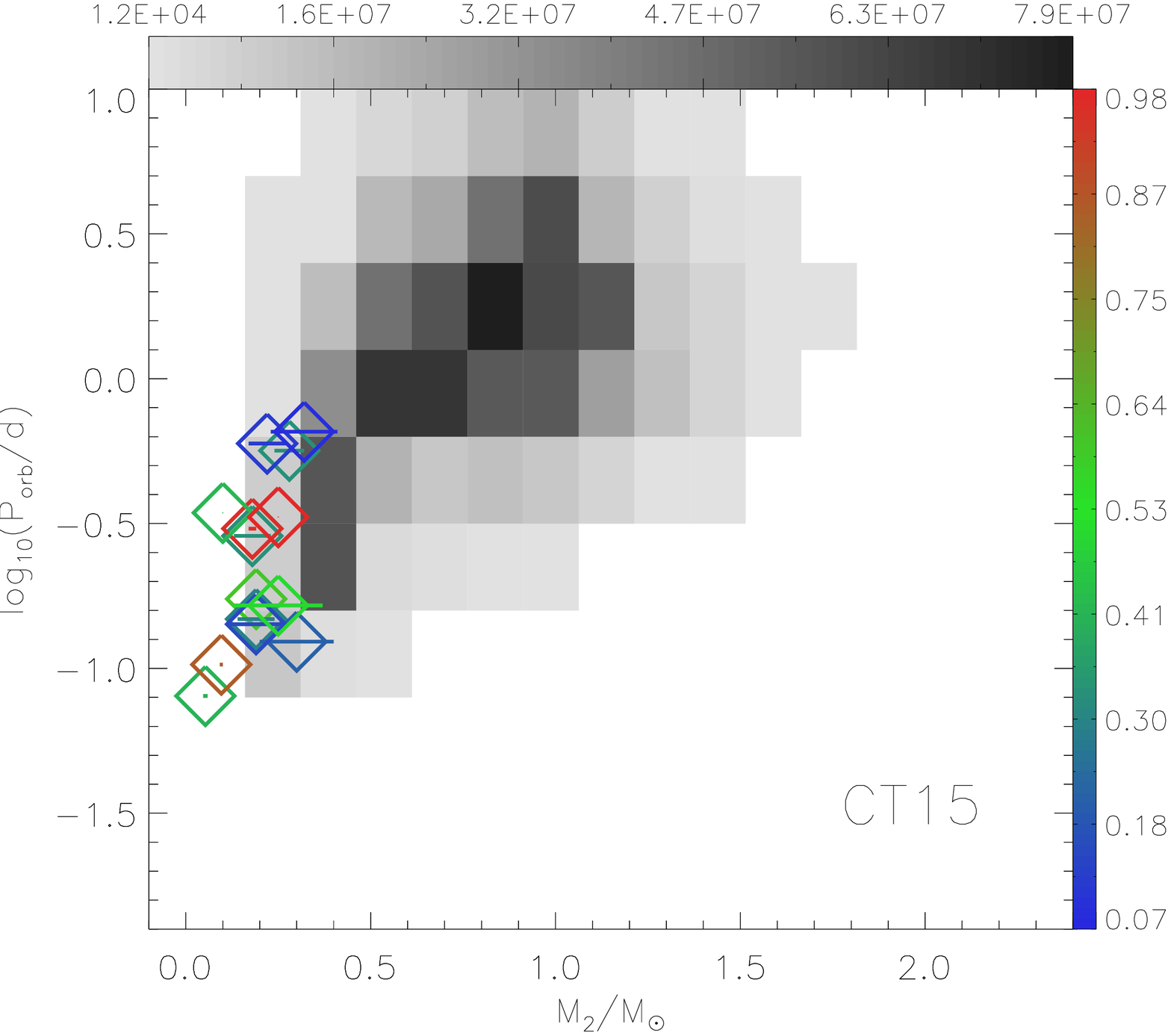}
    \caption{Same as for Fig. \ref{multiplot} but now showing the
      theoretical and observed PCEB population with
      $M_{\rmn{WD}}/\rmn{M}_{\odot}\le{0.4}$ for model PL2 (top panel)
      and for model CT15 (bottom panel); in both cases $n(q_{\rmn{i}})=1$. The
      grey-scale bar above the plots indicates the number of systems
      per bin area, while the bar on the right indicates the
      weightings of the observed systems.}
    \label{hPL2}
\end{figure}

As discussed above, the bulk of observed systems which have
$M_{2}/\rmn{M}_{\odot}\la{1.0}$ and $P_{\rmn{orb}}/\rmn{d}\la{1}$ are
consistent with our reference model A. IK Peg, on the other hand,
cannot be explained by models with $\alpha_{\rmn{CE}}\le{1.0}$, for
$\lambda=0.5$. Figure \ref{multiplot3} shows the theoretical PCEB
population with $1.1<M_{\rmn{WD}}/\rmn{M}_{\odot}\le{1.44}$ for a
range of models, indicated in the bottom right hand corner of each
panel. The location of IK Peg on the $M_{2}-\rmn{log}\,P_{\rmn{orb}}$
plane is indicated as the diamond in each case.

In order to explain the location of IK Peg, we require
$\alpha_{\rmn{CE}}\ga{3}$, if $\lambda=0.5$. Model PL2 achieves
$\alpha_{\rmn{CE}}\approx{2.9}$ for this system and, as shown in the
panel labelled `PL2' in Figure \ref{multiplot3} can account for the
location of IK Peg. However, this model also generates a low-mass
cut-off at $M_{2}\approx{0.4}$ M$_{\odot}$ in the PCEB
population. This is a feature which is not consistent with observed
systems, as highlighted in the top panel of Figure \ref{hPL2}. LM Com
and 0137-3457 have a 95 per cent and a 61 per cent probability of
sitting in this panel. Thus it appears this model cannot adequately
describe the CE phase in its present form.

Note that models CT0375 to CT15 cannot account for the location of IK
Peg either. From equation (\ref{alpha_func_mcut})
$\alpha_{\rmn{CE}}\rightarrow{1}$ as $M_{2}\rightarrow{\infty}$. As
with the model PL2, the cut-off (by design of the model) in the PCEB
population at $M_{2}=0.15$ M$_{\odot}$ doesn't appear to be supported
by the observed sample. The bottom panel of Figure \ref{hPL2} compares
the theoretical PCEB population with
$0.4<M_{\rmn{WD}}/\rmn{M}_{\odot}\le{0.5}$, with the corresponding
observed WD+MS systems. Indeed, model CT15 cannot account for the
observed location of, for example, HR Cam or 0137-3457, which have a
84 and 61 per cent probability of occupying this panel.

Model DTb, where the thermal energy of the giant's envelope
contributes to the ejection of the CE, can also account for the
location of IK Peg, while DTg cannot (see panels labelled `DTb' and
DTg' in Figure \ref{multiplot3}). Model DTg shows a slight increase in
the orbital period of the upper boundary in the theoretical PCEB
population, compared to that shown in the panel labelled `A', but this
is not enough to account for IK Peg. A possible primary
progenitor of IK Peg would fill its Roche lobe with a mass of 6
M$_{\odot}$ and a radius of 725 R$_{\odot}$. From Fig. \ref{lambda_R}
(dashed line), we find $\lambda_{\rmn{g}}\approx{0.6}$. However, to
account for the location of IK Peg, we require
$\lambda_{\rmn{g}}\approx{2.2}$ (with $\alpha_{\rmn{CE}}=1$).

Finally, the location of IK Peg can be accounted for if we consider an
angular momentum, rather than an energy, budget, as shown by the panel
`n15'. Of further interest is that, in contrast to the other models
considered in Figure \ref{multiplot4}, model n15 predicts \emph{no}
PCEBs with $M_{2}\la{1.0}$ M$_{\odot}$ at $P_{\rmn{orb}}\la{1}$
d. Furthermore, while the PCEB population in models A to DTb peaks at
$P_{\rmn{orb}}\approx{1}$ to 10 d, the model n15 population peaks at
$P_{\rmn{orb}}\approx{1000}$ d. This confirms the result by
\citet{maxted07}.

\subsection{The Initial Secondary mass Distribution}

\begin{figure*}
  \begin{minipage}{150mm}
    \centering
    \includegraphics[scale=0.1]{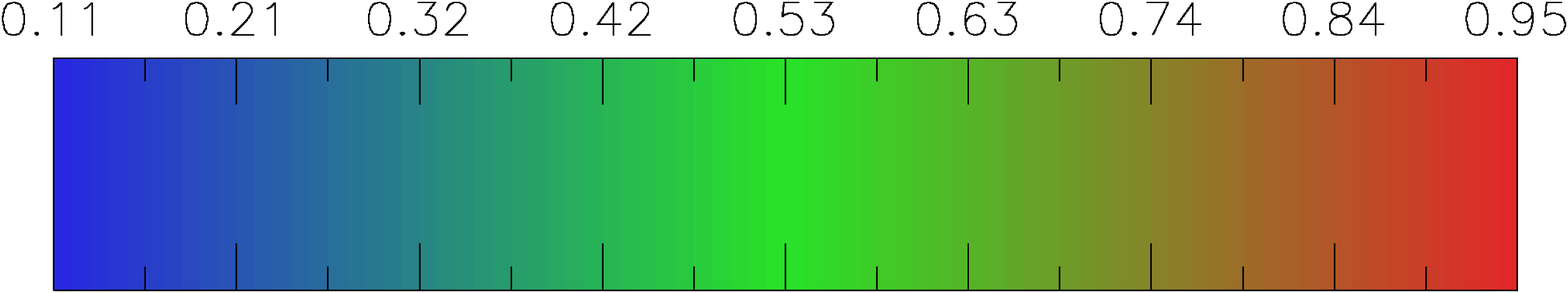}
    \includegraphics[scale=0.70]{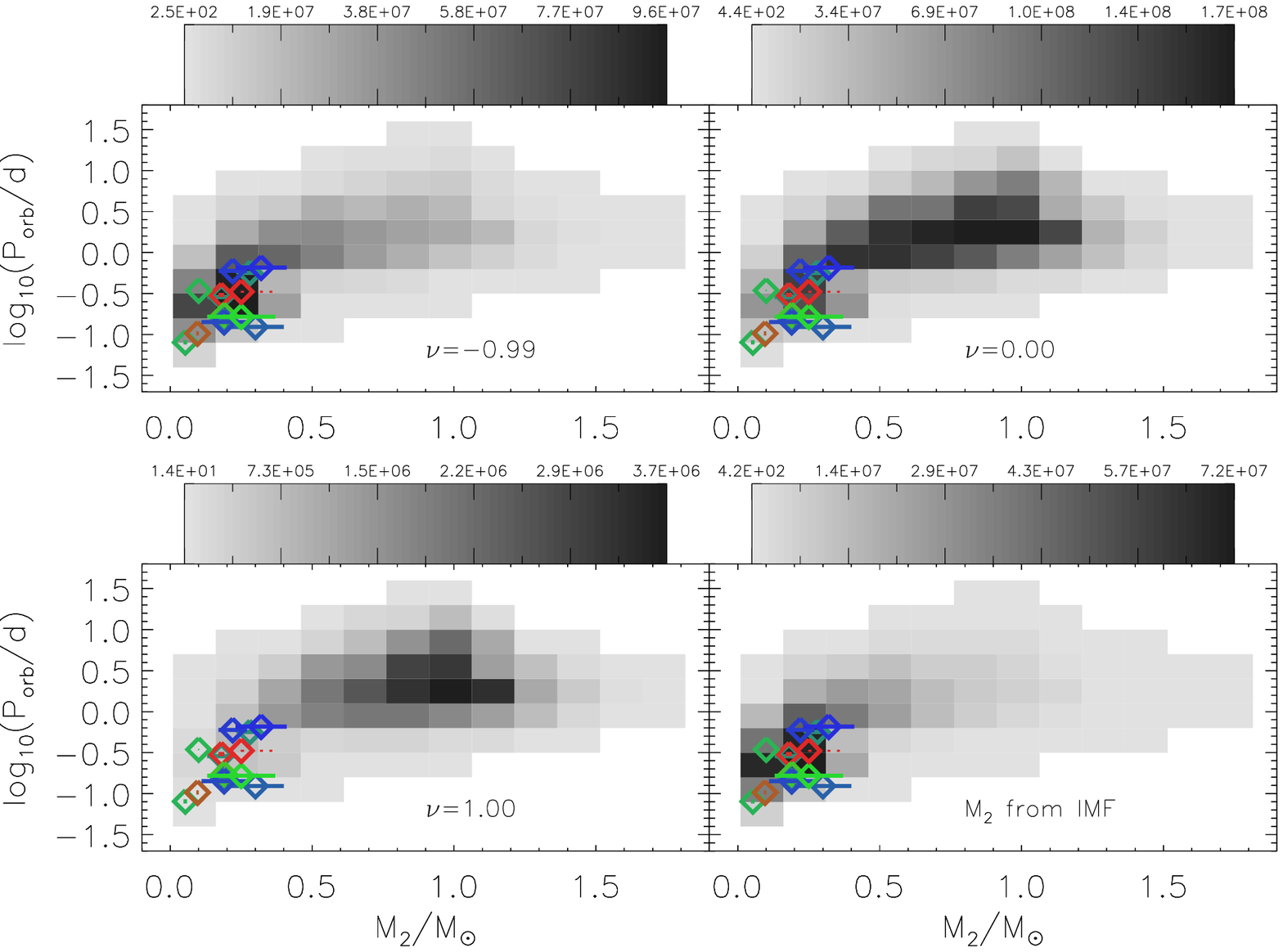}
    \caption{Same as Fig. \ref{multiplot}, but showing the population
    of PCEBs with $0.4<M_{\rmn{WD}}/\rmn{M}_{\odot}\le{0.5}$ for
    different initial secondary mass distributions, as indicated in
    each panel. The grey-scale bar above each panel gives the number
    of systems per bin area, while the colour bar at the top of the
    page gives the weighting of each system.}
    \label{multiplot4}
  \end{minipage}
\end{figure*}

We now consider the impact of the initial mass ratio distribution on
the theoretical PCEB populations. For our standard model A we
calculated the PCEB population for each initial distribution of the
secondary mass, which we compared to the observed WD+MS
systems. Figure \ref{multiplot4} illustrates the differences between
the resulting PCEB populations in the range
$0.4<M_{\rmn{WD}}/\rmn{M_{\odot}}\le{0.5}$ as an example.

For the cases $n(q_{\rmn{i}})\propto{q_{\rmn{i}}}^{-0.99}$ and IMFM2
the bulk of the PCEB population lies at $M_{2}\la{0.4}$ M$_{\odot}$
and $P_{\rmn{orb}}\la{1}$ d. This is consistent with the location of
the observed WD+MS systems in Figure \ref{multiplot4}. The theoretical
distribution with $n(q_{\rmn{i}})=1$ suggests a peak at
$M_{2}\approx{1}$ M$_{\odot}$ and $P_{\rmn{orb}}\approx{1}$ d, yet no
systems are observed in this region of
$M_{2}-\rmn{log}\,P_{\rmn{orb}}$ space. However, this is likely to be
a selection effect; PCEB candidates are selected according to their
blue colour due to the optical emission from the white dwarf and/or
high radial velocity variations. The flux from early-type secondaries
will dominate over that of the white dwarf, and hence these systems go
undetected. \citet{sg03} predicted that there is a large as yet
undetected population of old PCEBs, with cool white dwarfs and long
orbital period.

\citet{rebassa08} performed Monte-Carlo simulations to calculate the
detection probability of PCEBs with $P_{\rmn{orb}}/\rmn{d}\la{10}$,
based on the measurement accuracies of the Very Large Telescope
\citep{schreiber2008}, the SDSS, and for 1, 2 and $3\sigma$
significance of the radial velocity variations. They found that
approximately 6 out of their sample of 9 PCEBs should have
$P_{\rmn{orb}}>1$ d, yet this is not the case; all of their PCEBs have
$P_{\rmn{orb}}<1$ d, in contrast with the predictions of
\citet{sg03}. Thus it is possible that the sharp decline in the
population of PCEBs with $P_{\rmn{orb}}>1$ d is a characteristic of
the intrinsic PCEB population, making the models with
$n(q_{\rmn{i}})\propto{q_{\rmn{i}}}^{-0.99}$ or IMFM2 more
attractive. Note, however, that the local space density calculated for
an IMRD of $n(q_{\rmn{i}})\propto{q_{\rmn{i}}}^{-0.99}$ is in good
agreement with the observationally determined one, while IMFM2 is not.

\begin{figure}
  \includegraphics[scale=0.43]{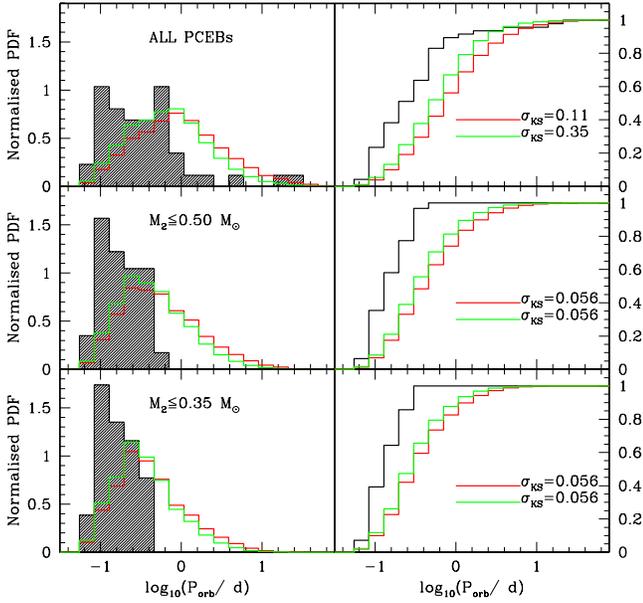}
  \caption{Left column: the normalised PDF of the orbital period
  distribution of the observed sample of PCEBs (hashed histogram)
  compared with the calculated distribution of the intrinsic
  population for $n(q_{\rmn{i}})=q_{\rmn{i}}^{-0.99}$ (red line),
  model A. The green line is similar to the red line, except here we
  also take into account the PCEB detection probability, as calculated
  by \citet{rebassa08}, assuming a measurement accuracy of 15 km
  s$^{-1}$ (appropriate for SDSS spectra), to detect 3$\sigma$ radial
  velocity variations. Top: all PCEBs; middle: PCEBs with
  $M_{2}\le{0.5}$ M$_{\odot}$; bottom: PCEBs with $M_{2}\le{0.35}$
  M$_{\odot}$. Right column: normalised CDFs of the corresponding
  orbital period distributions of the observed (solid black line),
  with the calculated distributions (red and green lines), with the
  scale indicated on the right axes. Also shown are the
  Kolmogorov-Smirnov significance levels, $\sigma_{\rmn{KS}}$, that
  the observed and calculated distributions are drawn from the same
  parent population.}
  \label{obs_theory_0.99}
\end{figure}

To determine if the intrinsic PCEB population does sharply decline for
$P_{\rmn{orb}}>1$ d, we compared the orbital period distribution of
our observed sample of PCEBs with our calculated distribution, with
$n(q_{\rmn{i}})=q_{\rmn{i}}^{-0.99}$, model A. These distributions are
shown as the hashed histogram and the red line in the top panel, left
column of Figure \ref{obs_theory_0.99}. The corresponding normalised
cumulative distribution functions (CDFs) are shown in the top panel,
right column of Figure \ref{obs_theory_0.99}, with the scale indicated
on the right axis. Note that, in contrast for the observed
distribution, the number of PCEBs in the intrinsic population
gradually declines for $P_{\rmn{orb}}\ga{1}$ d, as opposed to a sharp
decline.

We supplemented our calculated PCEB orbital period distribution with
the detection probabilities calculated by \citet{rebassa08} (see their
Figure 7). We considered the detection probabilities of WD+MS systems
showing radial velocity variations in their spectra with a $3\sigma$
significance (the criterion used by \citet{rebassa07} and
\citet{rebassa08} to identify PCEB candidates), as detected by the
SDSS (bottom curve of Fig. 7 in \citet{rebassa08}). As
\citet{rebassa08} only calculate the detection probabilities,
$\mathcal{P}$, for $P_{\rmn{orb}}\la{10}$ d, we extrapolated the curve
up to $P_{\rmn{orb}}=100$ d using the curve
$\mathcal{P}=0.43(P_{\rmn{orb}}/\rmn{d})^{-0.35}$.

The corresponding orbital period distribution is
shown as the green histogram in the top panels in the left and right
columns of Figure \ref{obs_theory_0.99}. Note that the inclusion of
the PCEB detection probabilities has a marginal effect. Indeed, we
still predict a gradual decline in the number of PCEBs with
$P_{\rmn{orb}}>1$ d.

To determine the likelihood that the observed and calculated PCEB
orbital period distributions are drawn from the same parent
distribution, we calculate the Kolmogorov-Smirnov statistic from the
normalised CDF distributions, and therefore the corresponding
significance level, $\sigma_{\rmn{KS}}$. A very small value of
$\sigma_{\rmn{KS}}$ shows that the two distributions are significantly
different, while $\sigma_{\rmn{KS}}=1$ shows that the two
distributions are in good agreement. We find $\sigma_{\rmn{KS}}=0.11$
when comparing the observed and calculated intrinsic (red) orbital
period distributions in the top panel, left column of Figure
\ref{obs_theory_0.99}. On the other hand, we find
$\sigma_{\rmn{KS}}=0.35$ between the observed and the calculated
(green) orbital period distribution with the detection probabilities
included.

We also consider the selection bias towards late-type secondaries by
only considering the observed and theoretical orbital period
distribution of PCEBs which have $M_{2}/\rmn{M}_{\odot}\le{0.5}$
(middle panels of Fig. \ref{obs_theory_0.99}) and
$M_{2}/\rmn{M}_{\odot}\le{0.35}$ (bottom panels). There is a better
agreement between the location of the peaks in the observed and
theoretical PCEB orbital period distributions. However, the
theoretical distributions (with and without the inclusion of PCEB
detection probabilities), still predict a gradual decline in the
number of PCEBs with $P_{\rmn{orb}}\ga{1}$ d, while there is a sharp
decline in the observed distribution. For the population of PCEBs with
$M_{2}\le{0.5}$ M$_{\odot}$ and M$_{\odot}\le{0.35}$, we find
$\sigma_{\rmn{KS}}=5.6\times{10}^{-2}$ between the observed and
calculated orbital period distributions, with and without the
inclusion of PCEB detection probabilities.

Thus, we cannot reproduce the observed sharp decline in the number of
PCEBs with $P_{\rmn{orb}}\la{1}$ d, even if we take into account in
our calculations the selection biases towards PCEBs with late-type
secondaries, and the biases against the detection of PCEBs with long
orbital periods. It is still unclear whether this sharp decline is
indeed a characteristic of the intrinsic PCEB population, or if it is
a result of further selection effects which have yet to be considered.

\section{Discussion}

\subsection{Constraining the CE phase}

We have shown that the majority of observed PCEBs (containing either a
sub-dwarf or white dwarf primary) can be reproduced by canonical
models with a constant, global value of $\alpha_{\rmn{CE}}>0.1$ for
the CE ejection efficiency. The systems V651 Mon, FF Aqr and V1379 Aql
are likely to have formed from a thermally unstable RLOF phase. This
is contrary to \citet{nt05} who assumed them to be PCEBs, and
attempted reconstruct their values of $\alpha_{\rmn{CE}}$. For the
case of V1379 Aql \citet{nt05} could not find a solution for
$\alpha_{\rmn{CE}}$, and hence took this system as evidence for their
`$\gamma$-algorithm'. We have shown that this system could have formed
from a thermally unstable RLOF phase.

There is only one system, IK Peg, that is both likely to be a PCEB and
at the same time inconsistent with the standard energy budget CE
model. Unlike the vast majority of the observed sample of PCEBs, this
system contains an early-type secondary star, and this may provide a
clue to the ejection mechanism during the CE phase.


Formally, the observed configuration of IK Peg requires
$\alpha_{\rmn{CE}}\ga{3}$. This means that a source of energy other
than gravitational potential energy is exploited for the ejection of
the CE. It has been suggested that this is the thermal and ionization
energy of the giant's envelope (Han et al. 1995, Han et al. 1995, Dewi
\& Tauris 2000, Webbink 2007). We find that by considering this extra
energy source in our models we can indeed account for the location of
IK Peg. However, this is a concept which has been challenged by
\citet{harpaz98} and \citet{sh03}. \citet{harpaz98} argue that, during
the planetary nebula phase, the opacity of the giant's envelope
decreases during recombination. Hence the envelope becomes transparent
to its own radiation. The radiation will therefore freely escape
rather than push against the material to eject it.

Previous population synthesis studies have considered constant, global
values of $\lambda=0.5$ (e.g. deKool 92, Willems \& Kolb
2004). \citet{dt00} suggested that this may lead to an overestimation
of the binding energy of the giant's envelope, and hence underestimate
the final PCEB orbital period. \citet{dt00} calculated values of
$\lambda_{\rmn{g}}$ for range of masses and radii, which we
incorporated into our population synthesis code. This, however, cannot
account for IK Peg either. We note that \citet{dt00} do not calculate
values of $\lambda_{\rmn{g}}$ for $R>600$ R$_{\odot}$. We concede that
we may still underestimate the value of $\lambda_{\rmn{g}}$ for the
case of IK Peg due to our adopted extrapolation to larger radii. If we
do linearly extrapolate $\lambda_{\rmn{g}}$ for a progenitor primary
of IK Peg with $M_{1}=6$ M$_{\odot}$ and $R_{1}=725$ R$_{\odot}$ we
find that $\lambda_{\rmn{g}}\approx{1.5}$, which is still not large
enough to account for IK Peg. It would therefore be beneficial to
calculate $\lambda$ in the mass and radius regime for PCEB
progenitors.

Rather than consider the CE phase in terms of energy,
\citet{nelemans00} and \citet{nt05} describe the CE phase in terms of
the angular momentum of the binary. Indeed, this is also a
prescription favoured by \citet{soker04}. Our model n15 can account
for the location of IK Peg. However, \citet{maxted07} found that for
the $\gamma$-prescription, the number of PCEBs increases with
increasing orbital periods larger than approximately 100 d. Indeed, we
also find increasing number of PCEBs with high mass white dwarfs with
increasing orbital period, with the bulk of the population lying at
approximately 1000 d. This is contrast with observations by
\citet{rebassa08}, who found a sharp decline in the number of PCEBs
with increasing orbital period beyond 1 d (however, see Section 4.2).

\subsection{Observing PCEBs}

Even though we have critically examined a variety of treatments for
the CE phase by comparing our models to the observed sample of PCEBs,
we have been unable to significantly constrain the underlying
physics. We believe this is mainly due to the selection effects still
pervading the observed sample of PCEBs. The large majority of our
observed PCEB sample contain late type secondaries, typically M3 to
M5. This is a consequence of the fact that until recently PCEB
candidates were identified in blue colour surveys, such as the Palomar
Green survey. As a result systems containing secondaries with early
spectral types will be missed, as their optical flux will dominate
over that from the white dwarf. A few exceptions include systems
identified by their large proper motions (e.g. RR Cae) or
spectroscopic binaries (e.g. V471 Tau). IK Peg was detected due to the
emission of soft X-rays from the young white dwarf, which has an
effective temperature of $40\:000$ K.

The present sample of PCEBs is therefore covering an insufficient
range in secondary masses. However, matters are improving with the
advent of the SDSS, which probes a large $ugriz$ colour space. This
will allow an extra 30 PCEBs to be supplemented to the currently known
46 systems in the foreseeable future (G\"{a}nsicke 2008, private
communication). A complementary program is currently underway to
target those WD+MS systems with cool white dwarfs and/or early-type
secondaries in order to compensate for the bias against such systems
in the previous surveys \citep{sns07}. It is therefore feasible that
we will be able to further constrain our models in the near future.

The observed sample of PCEBs also have $P_{\rmn{orb}}\la{1}$ d, which
can be argued to be a further selection effect; PCEBs are also
detected due to radial velocity variations of their spectra. However,
\citet{rebassa08} found that this long-period cut-off may be a
characteristic of the intrinsic PCEB population, rather than a
selection effect. More precisely, we have shown in Section 3.7 that
this may be a feature intrinsic to the population of PCEBs with
late-type secondaries.

We also find that an IMRD distribution of
$n(q_{\rmn{i}})\propto{q_{\rmn{i}}}^{-0.99}$ can reproduce the local
space density inferred from observed PCEBs by \citet{sg03}, as well as
accounting for the location of the currently known sample of PCEBs in
$M_{2}-\rmn{log}\:P_{\rmn{orb}}$ space. However, the more generally
preferred binary star IMRD is $n(q_{\rmn{i}})=1$ (e.g. Duquennoy \&
Mayor 1991, Mazeh et al. 1992, Goldberg, Mazeh \& Latham 2003).

\section{Conclusions}

By applying population synthesis techniques we have calculated the
present day population of post-common envelope binaries (PCEBs) for a
range of models describing the common envelope (CE) phase and for
different assumptions about the initial mass ratio distribution. We
have then compared these models to the currently known sample of PCEBs
in the three-dimensional configuration space made up of the two
component masses and the orbital period.

We find that the canonical model of a constant, global value of
$\alpha_{\rmn{CE}}>0.1$ can account for the observed PCEB systems with
late-type 
secondaries. However, this cannot explain IK Peg, which has
an early-type secondary star. 
IK Peg can be accounted for if we assume that the thermal
and ionization energy of the giant primary's envelope, as well as the
binary's orbital energy, can unbind the CE from the system. IK Peg can
also be explained by describing the CE phase in terms of the binary's
angular momentum, according to the $\gamma$-prescription proposed by
\citet{nelemans00} and \citet{nt05}.

We find that the present day number (local space density) of PCEBs in the 
Galaxy ranges from $3.0\times{10}^{5}$ ($6.0\times{10}^{-7}$ pc$^{-3}$) for
model CE01, $n(q_{\rmn{i}})\propto{q_{\rmn{i}}}^{-0.99}$, to
$1.1\times{10}^{9}$ ($2.2\times{10}^{-3}$ pc$^{-3}$) for model n15 and
for IMFM2.

We also find that an initial mass ratio distribution (IMRD) of
$n(q_{\rmn{i}})\propto{q_{\rmn{i}}}^{-0.99}$ gives local space
densities in the range
$6.0\times{10}^{-7}\la{\varrho}/\rmn{pc}^{-3}\la{5.6\times{10}^{-5}}$,
in good agreement with the observationally determined local space
density of
$6.0\times{10}^{-6}\la{\varrho}/\rmn{pc}^{-3}\la{3.0\times{10}^{-5}}$. This
form of the IMRD also predicts a decline in the population of PCEBs
with late-type ($M_{2}\le{0.35}$ M$_{\odot}$) secondaries, which is
what is observed by \citet{rebassa08}. However, while observations
show a sharp decline in the number of PCEBs with orbital periods
larger than 1 d, our theoretical calculations instead predict a
gradual decline. We cannot reproduce this sharp decline even if we
take into account observational biases towards PCEBs with late
spectral-type secondaries, and the selection biases against PCEBs with
orbital periods greater than 1 d.

Our work highlights that selection biases need to be overcome, especially for 
detecting PCEBs with early-type secondaries and/or cool white
dwarfs. This would greatly advance our understanding of the CE phase.

\section*{acknowledgements}

PJD acknowledges studentship support from the Science \& Technology
Facilities Council. We thank Boris G\"{a}nsicke for helpful comments
and data, and also Marc van der Sluys for the stellar evolution models
and useful discussion. Finally, we would like to thank the referee,
Christopher Tout, for his constructive comments which helped to
improve the presentation of the paper.


\begin{thebibliography}{}
  \bibitem[\protect\citeauthoryear{Aungwerojwit et al.}{2007}]{aungwerojwit07}%
    Aungwerojwit A., G\"{a}nsicke B.~T., Rodr{\'{\i}}guez-Gil P., Hagen H.~-J., %
    Giannakis )., Papadimitriou C., Allende Prieto C., Engels D., 2007, A\&A, %
    469, 297
  \bibitem[\protect\citeauthoryear{Bell et al.}{1994}]{bell94}%
    Bell S.~A., Pollacco D.~L., Hilditch R.~W., 1994, MNRAS, 270, 449
  \bibitem[\protect\citeauthoryear{Beer et al.}{2007}]{beer07}%
    Beer M.~E., Dray L.~M., King A.~R., Wynn G.~A., 2007, MNRAS, 375, 1000
  \bibitem[\protect\citeauthoryear{Bleach et al.}{2000}]{bleach00}%
    Bleach J.~N., Wood J.~H., Catal\'{a}n M.~S., Welsh W.~F., Robinson E.~L., %
    Skidmore W., 2000, MNRAS, 312, 70
  \bibitem[\protect\citeauthoryear{Bragaglia et al.}{1995}]{bragaglia95}%
    Bragaglia A., Renzini A., Bergeron P., 1995, ApJ, 443, 735
  \bibitem[\protect\citeauthoryear{Bruch}{1999}]{bruch99}%
    Bruch A., 1999, AJ, 117, 3031
  \bibitem[\protect\citeauthoryear{Catalan et al.}{1994}]{catalan94}%
    Catalan M.~S., Davey S.~C., Sarna M.~J., Cannon-Smith R., Wood J.~H., %
    1994, MNRAS, 269, 879
  \bibitem[\protect\citeauthoryear{Davis et al.}{2008}]{dkwg08}%
    Davis P.~J., Kolb U., Willems B., G\"{a}nsicke B.~T., 2008, MNRAS, 389, 1563
  \bibitem[\protect\citeauthoryear{de Kool}{1992}]{deKool92}%
    de Kool M., 1992, A\&A, 261, 188
  \bibitem[\protect\citeauthoryear{de Kool \& Ritter}{1993}]{dr93}%
    de Kool M., Ritter H., 1993, A\&A, 267, 397
  \bibitem[\protect\citeauthoryear{Dewi \& Tauris}{2000}]{dt00}%
    Dewi J. D. M., Tauris T. M., 2000, A\&A, 360, 1043
  \bibitem[\protect\citeauthoryear{Dewi \& Tauris}{2001}]{dt01}%
    Dewi J.~D.~M., Tauris T.~M., 2001, in Podsiadlowski P., Rappaport S., %
    King A.~R., D'Antona F., Burderi L., eds, ASPC 229, Evolution of Binary %
    and Multiple Star Systems, p. 255
  \bibitem[\protect\citeauthoryear{Drechsel et al.}{2001}]{drechsel01}%
    Drechsel H. et al., 2001, A\&A, 379, 893
  \bibitem[\protect\citeauthoryear{Duquennoy \& Mayor}{1991}]{dm91}%
    Duquennoy A., Mayor M., 1991, A\&A, 248, 285
  \bibitem[\protect\citeauthoryear{Etzel et al.}{1977}]{etzel77}%
    Etzel P.~B., Lanning H.~H., Patenaude D.~J., Dworetzki M.~M., %
    1977, PASP, 89, 616
  \bibitem[\protect\citeauthoryear{Ferguson et al.}{1999}]{ferguson99}%
    Ferguson D.~H., Liebert J., Haas S., Napiwotzki R., James T.~A., %
    1999, ApJ, 518, 866
  \bibitem[\protect\citeauthoryear{Fulbright et al.}{1993}]{fulbright93} %
    Fulbright M.~S., Liebert J., Bergeron P., Green R., 1993, ApJ, 406, 240
  \bibitem[\protect\citeauthoryear{G\"{a}nsicke et al.}{2004}]{gaensicke04}%
    G\"{a}nsicke B.~T., Araujo-Betancor S., Hagen H.~J., Harlaftis E.~T., Kitsionas S., %
    Dreizler S., Engels D., 2004, A\&A, 418, 265
  \bibitem[\protect\citeauthoryear{Goldberg, Mazeh \& Latham}{2003}]{gml03}%
    Goldberg D., Mazeh T., Latham D.~W., 2003, ApJ, 591, 397
  \bibitem[\protect\citeauthoryear{Good et al.}{2005}]{good05}%
    Good S.~A., Barstow M.~A., Burleigh M.~R., Dobbie P.~D., Holberg J.~B.,%
    2005, MNRAS, 364, 1082
  \bibitem[\protect\citeauthoryear{Green et al.}{1978}]{green78}%
    Green R.~F., Richstone D.~O., Schmidt M., 1978, ApJ, 224, 892
  \bibitem[\protect\citeauthoryear{Guinan \& Sion}{1984}]{gs84}%
    Guinan E.~F., Sion E.~M., 1984, AJ, 89, 1252
  \bibitem[\protect\citeauthoryear{Han et al.}{1994}]{han94}%
    Han Z., Podsiadlowski P., Eggleton P. P., 1994, MNRAS, 270, 121
  \bibitem[\protect\citeauthoryear{Han et al.}{1995}]{han95}%
    Han Z., Podsiadlowski P., Eggleton P. P., 1995, MNRAS, 272, 800
  \bibitem[\protect\citeauthoryear{Han et al.}{2002}]{han02}%
    Han Z., Podsiadlowski P., Maxted P.~F.~L., Marsh T.~R., Ivanova N., %
    2002, MNRAS, 336, 449
  \bibitem[\protect\citeauthoryear{Harpaz}{1998}]{harpaz98}%
    Harpaz A., 1998, ApJ, 498, 293
  \bibitem[\protect\citeauthoryear{Heber et al.}{2004}]{heber04}%
    Heber U. et al., 2004, A\&A, 420, 251
  \bibitem[\protect\citeauthoryear{Hilditch et al.}{1996}]{hilditch96}%
    Hilditch R.~W., Harries T.~J., Hill G., 1996, MNRAS, 279, 1380
  \bibitem[\protect\citeauthoryear{Hjellming \& Webbink}{1987}]{hw87}%
    Hjellming M. S., Webbink R. F., 1987, ApJ, 318, 794
  \bibitem[\protect\citeauthoryear{Hurley, Pols \& Tout}{2000}]{hpt00}%
    Hurley J. R., Pols O. R., Tout C. A., 2000, MNRAS, 315, 543
  \bibitem[\protect\citeauthoryear{Hurley, Tout \& Pols}{2002}]{htp02}%
    Hurley J. R., Tout C. A., Pols O.~R., 2002, MNRAS, 329, 897
  \bibitem[\protect\citeauthoryear{Iben \& Tutukov}{1984}]{it84}
    Iben I., Tutukov A. V., 1984, ApJS, 54, 235
  \bibitem[\protect\citeauthoryear{Iben \& Livio}{1993}]{il93}%
    Iben I. J., Livio M., 1993, PASP, 105, 1357
  \bibitem[\protect\citeauthoryear{Jeffery, Simon \& Evans}{1992}]{jse92}%
    Jeffery C.~S., Simon T., Evans T.~L., 1992, MNRAS, 258, 64
  \bibitem[\protect\citeauthoryear{Jeffery \& Simon}{1997}]{js97}%
    Jeffery C.~S., Simon T., 1997, MNRAS, 286, 487
  \bibitem[\protect\citeauthoryear{Kato, Nogami \& Baba}{2001}]{knb01}%
    Kato T., Nogami D., Baba H., 2001, PASJ, 53, 901
  \bibitem[\protect\citeauthoryear{Kawka et al.}{2002}]{kawka02}%
    Kawka A., Vennes S., Koch R., Williams A., 2002, AJ, 124, 2853
  \bibitem[\protect\citeauthoryear{Kawka et al.}{2008}]{kawka08}%
    Kawka A., Vennes S., Dupuis J., Chayer P., Lanz T., 2008, ApJ, 675, 1518
  \bibitem[\protect\citeauthoryear{Kilkenny et al.}{1998}]{kilkenny98}%
    Kilkenny D., O'Donoghue D., Koen C., Lynas-Gray A.~E., van Wyk F., 1998, %
    MNRAS, 296, 329
  \bibitem[\protect\citeauthoryear{Kroupa, Tout \& Gilmore}{1993}]{ktg93}%
    Kroupa P., Tout C. A., Gilmore G., 1993, MNRAS, 262, 545
  \bibitem[\protect\citeauthoryear{Landsman, Simon \& Bergeron}{1993}]{lsb93}%
    Landsman W., Simon T., Bergeron P., 1993, PASP, 105, 841
  \bibitem[\protect\citeauthoryear{Lanning \& Pesch}{1981}]{lp81}%
    Lanning H.~H., Pesch P., 1981, ApJ, 244, 280
  \bibitem[\protect\citeauthoryear{Marsh \& Duck}{1996}]{md96}%
    Marsh T.~R., Duck S.~R., 1996, MNRAS, 278, 565
  \bibitem[\protect\citeauthoryear{Maxted et al.}{1998}]{maxted98}%
    Maxted P.~F.~L., Marsh T.~R., Moran C., Dhillon V.~S., Hilditch R.~W., %
    1998, MNRAS, 300, 1225
  \bibitem[\protect\citeauthoryear{Maxted et al.}{2002}]{maxted02}%
    Maxted P. F. L., Burleigh M. R., Marsh T. R., Bannister N. P., %
    2002, MNRAS, 334, 833
  \bibitem[\protect\citeauthoryear{Maxted et al.}{2002b}]{maxted02b}%
    Maxted P.~F.~L., Marsh T.~R., Heber U., Morales-Rueda L., North R.~C., %
    Lawson W.~A., 2002, MNRAS, 333, 231
  \bibitem[\protect\citeauthoryear{Maxted et al.}{2004}]{maxted04}%
    Maxted P.~F.~L., Marsh T.~R., Morales-Rueda L., Barstow M.~A., %
    Dobbie P.~D., Schreiber M.~R., Dhillon V.~S., Brinkworth C.~S., %
    2004, MNRAS, 355, 1143
  \bibitem[\protect\citeauthoryear{Maxted et al.}{2006}]{maxted06}%
    Maxted P.~F.~L., Napiwotzki R., Dobbie P.~D., Burleigh M.~R., %
    2006, Nature, 442, 543
  \bibitem[\protect\citeauthoryear{Maxted et al.}{2007}]{maxted07}%
    Maxted P.~F.~L., Napiwotzki R., Marsh T.~R., Burleigh M.~R., %
    Dobbie P.~D., Hogan E., Nelemans G., 2007, in Napiwotzki R., %
    Burleigh M.~R., eds., ASP Conf. Ser., vol. 372, 15th. European %
    Workshop on White Dwarfs, p. 471
  \bibitem[\protect\citeauthoryear{Mazeh et al.}{1992}]{mazeh92}%
    Mazeh T., Goldberg D., Duquennoy A., Mayor M., 1992, ApJ, 401, 265
  \bibitem[\protect\citeauthoryear{Mendez \& Niemela}{1981}]{mn81}%
    Mendez R.~H., Niemela V.~S., 1981, ApJ, 250, 240
  \bibitem[\protect\citeauthoryear{Mendez et al.}{1985}]{mendez85}%
    Mendez R.~H., Marino R.~F., Claria J.~J., van Driel W., 1985, %
    Rev. Mex. Astron. Astrofis., 10, 187
  \bibitem[\protect\citeauthoryear{Miller et al.}{1976}]{miller76}%
    Miller J.~S., Krzeminski W., Priedhorsky W., 1976, %
    IAU Circ., 2974, 1
  \bibitem[\protect\citeauthoryear{Morales-Rueda et al.}{2005}]{morales05}%
    Morales-Rueda L., Marsh T.~R., Maxted P.~F.~L., Nelemans G., Karl C., %
    Napiwotzki R., Moran C.~J.~K., 2005, MNRAS, 359, 648
  \bibitem[\protect\citeauthoryear{Nelemans et al.}{2000}]{nelemans00}%
    Nelemans G., Verbunt F., Yungelson L. R., Portegies Zwart S. F., 2000,%
    A\&A,360,1011
  \bibitem[\protect\citeauthoryear{Nelemans \& Tout}{2005}]{nt05}%
    Nelemans G., Tout C. A., 2005, MNRAS, 356, 753
  \bibitem[\protect\citeauthoryear{Nemeth et al.}{2005}]{nemeth05}%
    Nemeth P., Kiss L.~L., Sarneczky K., 2005, IBVS, 5599, 8
  \bibitem[\protect\citeauthoryear{O'Brien et al.}{2001}]{obrien01}%
    O'Brien M.~S., Bond H.~E., Sion E.~M., 2001, ApJ, 563, 971
  \bibitem[\protect\citeauthoryear{O'Donoghue et al.}{2003}]{odonoghue03}%
    O'Donoghue D., Koen C., Kilkenny D., Stobie R.~S., Koester D., Bessell M.~S., %
    Hambley N., MacGillivrayH., 2003, MNRAS, 345, 406
  \bibitem[\protect\citeauthoryear{{\O}stensen et al.}{2007}]{ostensen07}%
    {\O}stensen R., Oreiro R., Drechsel H., Heber U., Baran A., Pigulski A., %
    2007, in Napiwotzki R., Burleigh M.~R., eds, ASP Conf. Ser., Vol. 372, %
    15th. European Workshop on White Dwarfs, p. 483
  \bibitem[\protect\citeauthoryear{Paczy\'{n}ski \& Zi{\'o}{\l}kowski}{1967}]{pz67}%
    Paczy{\'n}ski B., Zi{\'o}{\l}kowski J., 1967, AcA, 17, 7
  \bibitem[\protect\citeauthoryear{Paczy\'{n}ski}{1971}]{paczynski71}%
    Paczy{\'n}ski B., 1971, ARA\&A, 9, 183
  \bibitem[\protect\citeauthoryear{Paczy\'{n}ski}{1976}]{paczynski76}%
    Paczy\'{n}ski P., 1976, in Eggleton P., Mitton S., Whelan J., eds,
    Proc. IAU Symp. 73, Structure and Evolution of Close Binary
    Systems, p. 75
  \bibitem[\protect\citeauthoryear{Paxton}{2004}]{paxton04}%
    Paxton B., 2004, PASP, 116, 699
  \bibitem[\protect\citeauthoryear{Politano}{2004}]{politano04}%
    Politano M., 2004, ApJ, 604, 817
  \bibitem[\protect\citeauthoryear{Politano \& Weiler}{2007}]{pw07}%
    Politano M., Weiler M., 2007, ApJ, 665, 663
  \bibitem[\protect\citeauthoryear{Pollacco \& Bell}{1993}]{pb93}%
    Pollacco D.~L., Bell S.~A., 1993, MNRAS, 262, 377
  \bibitem[\protect\citeauthoryear{Polubek et al.}{2007}]{polubek07}%
    Polubek G., Pigulski A., Baran A., Udalski A., 2007, in %
    Napiwotzki R., Burleigh M.~R., eds, ASP Conf. Ser., Vol. 372, %
    15th. European Workshop on White Dwarfs, p. 487
  \bibitem[\protect\citeauthoryear{Rappaport, Verbunt \& Joss}{1993}]{rvj83}
    Rappaport S., Verbunt F., Joss P.~C., 1983, ApJ, 275, 713
  \bibitem[\protect\citeauthoryear{Rauch}{2000}]{rauch00}%
    Rauch T., 2000, A\&A, 356, 665
  \bibitem[\protect\citeauthoryear{Rebassa-Mansergas et al.}{2007}]{rebassa07}%
    Rebassa-Mansergas A., G\"{a}nsicke B.~T., Rodr{\'{\i}}guez-Gil P., Schreiber M.~R.,%
    Koester D., 2007, MNRAS, 382, 1377
  \bibitem[\protect\citeauthoryear{Rebassa-Mansergas et al.}{2008}]{rebassa08}%
    Rebassas-Mansergas A., G\"{a}nsicke B.~T., Schreiber M.~R., Southworth J., %
    Schwope A.~D., Nebot Gomez-Moran A., Aungwerojwit A., Rodr{\'{\i}}guez-Gil P., %
    Karamanavis V., Krumpe M., Tremou E., Schwarz R., Staude A., Vogel J., 2008, MNRAS
  \bibitem[\protect\citeauthoryear{Ritter \& Kolb}{2003}]{rk03}%
    Ritter H., Kolb U., 2003, A\&A, 404, 301
  \bibitem[\protect\citeauthoryear{Saffer et al.}{1993}]{saffer93}%
    Saffer R.~A. et al., 1993, AJ, 105, 1945
  \bibitem[\protect\citeauthoryear{Sandquist, Taam \& Burkert}{2000}]{stb00}%
    Sandquist E. L., Taam R. E., Burkert A., 2000, ApJ, 533, 948
  \bibitem[\protect\citeauthoryear{Schmidt et al.}{1995}]{schmidt95}%
    Schmidt G.~D., Smith P.~S., Harvey D.~A., Grauer A.~D., 1995, AJ, 110, 398
  \bibitem[\protect\citeauthoryear{Schreiber \& G\"{a}nsicke}{2003}]{sg03}%
    Schreiber M.~R., G\"{a}nsicke B.~T., 2003, A\&A, 406, 305
  \bibitem[\protect\citeauthoryear{Schreiber, Nebot Gomez-Moran \& Schwope}{2007}]{sns07}%
    Schreiber M.~R., Nebot Gomez-Moran A., Schwope A., 2007, in Napiwotzki R., Burleigh M.~R., %
    eds., 15th. European Workshop on White Dwarfs, ASP Conf. Ser., 372, 459
  \bibitem[\protect\citeauthoryear{Schreiber et al.}{2008}]{schreiber2008}%
    Schreiber M.~R., {G{\"a}nsicke} B.~T., Southworth J., Schwope A.~D., %
    Koester D., 2008, MNRAS, 484, 441
  \bibitem[\protect\citeauthoryear{Shimanskii et al.}{2004}]{shimanskii04}%
    Shimanskii V.~V., Borisov N.~V., Sakhibullin N.~A., Surkov A.~E., 2004, %
    Astron. Rep., 48, 563
  \bibitem[\protect\citeauthoryear{Shimansky et al.}{2003}]{shimansky03}%
    Shimansky V.~V., Borisov N.~V., Shimanskaya N.~N., 2003, Astron. Rep., %
    47, 763
  \bibitem[\protect\citeauthoryear{Smalley et al.}{1996}]{smalley96}
    Smalley B., Smith K.~C., Wonnacott D., Allen C.~S., 1996, MNRAS, 278, 688
  \bibitem[\protect\citeauthoryear{Soker}{2004}]{soker04}%
    Soker N., in Tovmassian G., Sion E., eds., Revista Mexicana de Astronomia y %
    Astrofisica Conf. Ser., Vol. 20, p. 30
  \bibitem[\protect\citeauthoryear{Soker \& Harpaz}{2003}]{sh03}%
    Soker N., Harpaz A., 2003, MNRAS, 343, 456
  \bibitem[\protect\citeauthoryear{Stauffer}{1987}]{stauffer87}%
    Stauffer J.~R., 1987, AJ, 94, 996
  \bibitem[\protect\citeauthoryear{Taam \& Sandquist}{2000}]{ts00}%
    Taam R. E., Sandquist E. L., 2000, ARA\&A, 38, 113
  \bibitem[\protect\citeauthoryear{Tappert et al.}{2007}]{tappert07}%
    Tappert C., G\"{a}nsicke B.~T., Schmidtobreick L., Aungwerojwit A., %
    Menneickent R.~E., Koester D., 2007, A\$A, 474, 205
  \bibitem[\protect\citeauthoryear{Tout \& Eggleton}{1988}]{te88}%
    Tout C. A., Eggleton P. P., 1988, MNRAS, 231, 823
  \bibitem[\protect\citeauthoryear{Tout \& Eggleton}{1988b}]{te88b}%
    Tout C.~A., Eggleton P.~P., 1988b, ApJ, 334, 357
  \bibitem[\protect\citeauthoryear{Unglaub}{2008}]{unglaub08}%
    Unglaub K., 2008, A\&A, 486, 923
  \bibitem[\protect\citeauthoryear{van den Besselaar}{2007}]{besselaar07}%
    van den Besselaar E.~J.~M. et al., 2007, A\&A, 466, 1031
  \bibitem[\protect\citeauthoryear{Vaccaro \& Wilson}{2003}]{vw03}%
    Vaccaro T.~R., Wilson R. E., 2003, MNRAS, 342, 564
  \bibitem[\protect\citeauthoryear{Vennes \& Thorstensen}{1994}]{vt94}%
    Vennes S., Thorstensen J.~R., 1994, AJ, 108, 1881
  \bibitem[\protect\citeauthoryear{Vennes, Christian \& Thorstensen}{1998}]{vct98}%
    Vennes S., Christian D.~J., Thorstensen J.~R., 1998, ApJ, 502, 763
  \bibitem[\protect\citeauthoryear{Vennes et al.}{1999}]{vennes99}%
    Vennes S., Thorstensen J.~R., Polomski E.~F., 1999, ApJ, 523, 386
  \bibitem[\protect\citeauthoryear{Vink}{2004}]{vink04}%
    Vink J.~S., 2004, Ap\&SS, 291, 239
  \bibitem[\protect\citeauthoryear{Webbink}{2007}]{webbink07}%
    Webbink R. F., in Milone E. F., Leahy D. A., Hobill D. W., eds, %
    Short Period Binary Stars, Springer
  \bibitem[\protect\citeauthoryear{Weidemann}{1990}]{weidemann90}%
    Weidemann V., 1990, ARA\&A, 28, 103
  \bibitem[\protect\citeauthoryear{Willems \& Kolb}{2002}]{wk02}%
    Willems B., Kolb U., 2002, MNRAS, 337, 1004
  \bibitem[\protect\citeauthoryear{Willems \& Kolb}{2004}]{wk04}%
    Willems B., Kolb U., 2004, A\&A, 419, 1057
  \bibitem[\protect\citeauthoryear{Willems et al.}{2005}]{willems05}%
    Willems B., Kolb U., Sandquist E. L., Taam R. E., Dubus G., 2005, ApJ, %
    635, 1263
  \bibitem[\protect\citeauthoryear{Wils et al.}{2007}]{wils07}%
    Wils P., Di Scala G., Otero S.~A., 2007, IBVS No. 5800
  \bibitem[\protect\citeauthoryear{Wood et al.}{1995}]{wood95}%
    Wood J.~H., Robinson E.~L., Zhang E.~-H., 1995, MNRAS, 277, 87
  \bibitem[\protect\citeauthoryear{Wood \& Saffer}{1999}]{ws99}%
    Wood J.~H., Saffer R., 1999, MNRAS, 305, 820
\end{thebibliography}
\end{document}